\documentclass[journal]{IEEEtran}
\usepackage{amsmath,amsfonts,amssymb}

\usepackage{algorithm}
\usepackage{array}
\usepackage{algpseudocode}

\usepackage[caption=false,font=normalsize,labelfont=sf,textfont=sf]{subfig}
\usepackage{textcomp}
\usepackage{stfloats}
\usepackage{url}
\usepackage{verbatim}
\usepackage{graphicx}
\usepackage{cite}
\usepackage{textcomp}
\usepackage{tabularx}
\usepackage{stfloats}
\usepackage{url}
\usepackage{makecell}
\usepackage{balance}
\usepackage{color}
\usepackage{enumerate}
\usepackage{orcidlink} 
\usepackage{booktabs}
\usepackage[normalem]{ulem}
\usepackage{xcolor}
\usepackage{hyperref}
\hypersetup{
	colorlinks=true,
	linkcolor=blue,
	citecolor=blue,
}
\usepackage{latexsym,bm}
\usepackage{amsmath,amssymb}
\usepackage{amsfonts,amssymb}

\begin{document}
\title{\LARGE 220 GHz RIS-Aided Multi-user Terahertz Communication System: Prototype Design and Over-the-Air Experimental Trials}

\author{Yanzhao Hou$^{\orcidlink{0000-0001-5571-9539}}$,~\IEEEmembership{Member,~IEEE}, Guoning Wang$^{\orcidlink{0009-0009-4623-9109}}$, Chen Chen$^{\orcidlink{0009-0004-6536-3002}}$, Gaoze~Mu$^{\orcidlink{0000-0002-2881-9902}}$,~\IEEEmembership{Member,~IEEE},\\ Qimei Cui$^{\orcidlink{0000-0003-1720-220X}}$,~\IEEEmembership{Senior Member,~IEEE}, Xiaofeng~Tao$^{\orcidlink{0000-0001-9518-1622}}$,~\IEEEmembership{Senior Member,~IEEE}, Yuanmu Yang$^{\orcidlink{0000-0002-5264-0822}}$\vspace{-1.5em}
        % <-this % stops a space
% <-this % stops a space
%\thanks{Manuscript received April 19, 2021; revised August 16, 2021.}
\thanks{This work was supported in part by the National Natural Science Foundation of China under 62327801, in part by the Joint Research Fund for Beijing Natural Science Foundation and Haidian Original Innovation under Grant L232001, and in part by the 111 Project of China (No. B16006). \textit{(Corresponding author: Yuanmu Yang.)} }
\thanks{Yanzhao Hou, Qimei Cui, and Xiaofeng Tao are with the National Engineering Laboratory for Mobile Networks, Beijing University of Posts and Telecommunications, Beijing 100876, China, and also with Pengcheng Laboratory, Shenzhen 518055, China. (e-mail: houyanzhao@bupt.edu.cn; 
cuiqimei@bupt.edu.cn; taoxf@bupt.edu.cn).}
\thanks{Guoning Wang and Gaoze Mu are with the School of Information and Communication Engineering, Beijing University of Posts and Telecommunications, Beijing 100876, China. (e-mail: wguoning@bupt.edu.cn; mugz@bupt.edu.cn).}
\thanks{Chen Chen and Yuanmu Yang are with the State Key Laboratory for Precision Measurement Technology and Instruments, Department of Precision Instrument, Tsinghua University, Beijing 100084, China. (e-mail: cchen22@mails.tsinghua.edu.cn; ymyang@tsinghua.edu.cn).}
\thanks{These authors contributed equally: Yanzhao Hou, Guoning Wang, and Chen Chen.}
}

%  \markboth{Journal of \LaTeX\ Class Files,~Vol.~14, No.~8, August~2021}%
%  {Shell \MakeLowercase{\textit{et al.}}: A Sample Article Using IEEEtran.cls for IEEE Journals}

\IEEEpubid{}

\maketitle

\begin{abstract}
Terahertz (THz) communication technology is regarded as a promising enabler for achieving ultra-high data rate transmission in next-generation communication systems. To mitigate the high path loss in THz systems, the transmitting beams are typically narrow and highly directional, which makes it difficult for a single beam to serve multiple users simultaneously. To address this challenge, reconfigurable intelligent surfaces (RIS), which can dynamically manipulate the wireless propagation environment, have been integrated into THz communication systems to extend coverage. Existing works mostly remain theoretical analysis and simulation, while prototype validation of RIS-assisted THz communication systems is scarce. In this paper, we designed a liquid crystal-based RIS operating at 220 GHz supporting both single-user and multi-user communication scenarios, followed by a RIS-aided THz communication system prototype. To enhance the system performance, we developed a beamforming method including a real-time power feedback control, which is compatible with both single-beam and multi-beam modes. To support simultaneous multi-user transmission, we designed an OFDM-based resource allocation scheme. In our experiments, the received power gain with RIS is no less than 10 dB in the single-beam mode, and no less than 5 dB in the multi-beam mode. With the assistance of RIS, the achievable rate of the system could reach 2.341 Gbps with 3 users sharing 400 MHz bandwidth and the bit error rate (BER) of the system decreased sharply. Finally, an image transmission experiment was conducted to vividly show that the receiver could recover the transmitted information correctly with the help of RIS. The experimental results also demonstrated that the received signal quality was enhanced through power feedback adjustments.
\end{abstract}


\vspace{-0.5em}
\section{Introduction}
\IEEEPARstart{W}{ith} the widespread deployment of the fifth-generation (5G) communication networks, discussions about next-generation communication technologies have begun to emerge. The sixth-generation (6G) communication system presents new demands for achievable rates, reliability, and connectivity \cite{saad2019vision,akyildiz20206g,yang20196g}. Existing research indicates that by 2030, the peak rates supported by 6G networks are expected to reach terabits per second (Tbps) level, enabling connectivity anytime and anywhere \cite{ji2021several,chafii2023twelve,yuan2020potential}. Terahertz (THz) communication technology, recognized for its wide bandwidth, high transmission speed, and enhanced security, has been identified as one of the key enabling technologies for 6G \cite{shafie2022terahertz,alsharif2021toward,chowdhury20206g,yang2022terahertz}. However, due to the high-frequency characteristics, THz waves often suffer from significant free-space path loss and are easily obstructed by obstacles, resulting in a limited coverage area \cite{10255735,song2021terahertz}. On the other hand, in 5G systems, large-scale multiple-input multiple-output (MIMO) is employed to achieve precise beamforming and provide spatial multiplexing for multiple users \cite{chataut2020massive}. As frequency increases, large-scale MIMO operating at THz band requires integrating more antenna elements within a unit volume, leading to high power consumption and complexity, making it challenging to implement in the short term \cite{do2021terahertz}. Therefore, there is an urgent need for new enabling technologies to expand the coverage of THz communication systems while maintaining lower power consumption.

Reconfigurable intelligent surface (RIS), also known as intelligent reflecting surface (IRS) or reconfigurable metasurface, is a planar surface composed of a large number of passive reflecting elements \cite{liu2021reconfigurable,yuan2021reconfigurable}. By jointly designing the phase response of each element, flexible control of incident electromagnetic waves can be achieved with extremely low power consumption \cite{pan2021reconfigurable,basharat2021reconfigurable}. Due to such an advantage, RIS has been considered a potential technology in 6G. Integrating RIS into THz communication systems allows for the redirection of THz beams, extending the coverage and enhancing achievable rates \cite{10720781,raza2022intelligent,yang2022terahertz2}. The authors in \cite{huang2021multi} proposed a deep reinforcement learning (DRL) based method to jointly design the digital beamforming at base station (BS) and analog beamforming at RIS, and then proved that the coverage of existing THz communication systems could be improved by $50\%$. In \cite{do2024throughput}, a joint optimization of power allocation at BS and reflection coefficients at RIS is performed and the sum rate of the investigated RIS-assisted THz-MIMO system is significantly improved.

\subsection{Related Works}\IEEEpubidadjcol 

\begin{table*}[t]
	\footnotesize
	%	\vspace{-0.5cm}
	\caption{summary of ris prototype related works}\vspace{-1em}
	\begin{center}
		\begin{tabular}{c|l|l|l|m{2.2cm}|m{2.4cm}|m{1.8cm}|l}
			\Xhline{0.7px}
			\textbf{Reference} &\textbf{Frequency} &\textbf{Bandwidth} & \textbf{Material} & \textbf{Function of RIS} & \textbf{Data Transmission} & \textbf{Beam Number }& \textbf{Peak Rates} \\
                \hline
			  \cite{dai2019wireless} & 4 GHz & Around 0.5 GHz & Diode &\multicolumn{1}{|m{2.1cm}|}{Modulation} & Yes, QPSK & Single beam & 1638.4 kbps\\
                \hline
			\cite{liu2023toward} & 88 GHz & Around 14 GHz & Diode & \multicolumn{1}{|m{2.1cm}|}{Modulation} & Yes, FSK,QPSK,16QAM & Single beam & 100 Mbps\\
                \hline
			\cite{wu2023universal} & 23.5 GHz & Unmentioned & Diode & \multicolumn{1}{|m{2.1cm}|}{Modulation} & Yes, QPSK,QAM &  Single beam & 2 Mbps \\
                \hline
			\cite{wang2022broadband}  & 218-232 GHz & 14 GHz & $\mathrm{VO_{2}}$ & \multicolumn{1}{|m{2.1cm}|}{Beam Steering (range: $>50^\circ$)} & No, single-tone only & Single beam &-\\
                \hline
			\cite{lv2023broadband} & 735-965 GHz & 230 GHz & $\mathrm{VO_{2}}$ & \multicolumn{1}{|m{2.1cm}|}{Beam Steering (range: $>50^\circ$)} & No, single-tone only & Single beam &-\\
                \hline
                \cite{wu2020liquid} & 683 GHz & Around 50 GHz & Liquid Crystal & \multicolumn{1}{|m{2.1cm}|}{Beam Steering (range: $32^\circ$)} & No, single-tone only & Single beam &-\\
                \hline
                \cite{li2023modulo} & 728 GHz & Around 15 GHz & Liquid Crystal & \multicolumn{1}{|m{2.1cm}|}{Beam Steering (2-dimensional)} & No, single-tone only & Single beam &-\\
                \hline
                \cite{chen2024liquid} & 290 GHz & Around 23 GHz & Liquid Crystal & \multicolumn{1}{|m{2.1cm}|}{Beam Steering (range: $110^\circ$)}& No, single-tone only &  3 beams simultaneously &-\\
                \hline
                Our Work & 220 GHz & Around 15 GHz & Liquid Crystal &\multicolumn{1}{|m{2.1cm}|}{Beam Steering (range: $110^\circ$)} & Yes, QPSK,16QAM &  3 beams simultaneously & 2.341 Gbps \\
			\Xhline{0.7px}
		\end{tabular}
	\end{center}\label{tab0}\vspace{0em}
\end{table*}

So far, with the ongoing advancement of electromagnetic meta-material technology, research on the prototype design of RIS-assisted communication systems is flourishing. Related works are summarized in Tab. \ref{tab0}. Based on the roles that RIS plays in communication systems, existing works can be roughly divided into two categories: modulating the incident signal with RIS, or performing beam steering with RIS. 

In the first category, the authors in \cite{dai2019wireless} developed a RIS based on diodes. When a time-varying bias voltage is applied, the designed RIS could perform QPSK modulation on a 4 GHz carrier at a maximum rate of 1638.4 kbps. In \cite{liu2023toward}, a more complex control mechanism for RIS is introduced, allowing simultaneous adjustment of the amplitude and phase of the incident wave by varying the amplitude and duty cycle of the bias voltage. This system operates at a center frequency of 88 GHz and supports FSK, QPSK, and 16QAM modulation modes, achieving a maximum rate of up to 100 Mbps. Similarly, the authors in \cite{wu2023universal} designed a RIS operating at 23.5 GHz and developed a space-time coding scheme that supports QPSK and QAM modulation modes, achieving a maximum rate of 2 Mbps. 

In these researches, the primary intention of introducing RIS is to reduce the complexity of radio frequency (RF) frontend in microwave and millimeter-wave transmitters. However, in these systems, the modulation rate is largely constrained by the switching speed of the RIS. On the other hand, diodes exhibit a certain characteristic of frequency selectivity meaning traditional lumped diode-based RIS typically cannot operate in the THz band \cite{10177872}. Instead, the THz-band RIS is generally constructed from tunable materials such as liquid crystals and vanadium dioxide ($\mathrm{VO_{2}}$)\cite{liu2018programmable}. The response time of them is generally much greater than that of diodes, which makes it challenging for THz-band RIS to modulate incoming signals at high rates. Therefore, RIS in the THz band is typically intended to redirect highly directional THz beams rather than modulate the incident signal.

In the second category, the authors in \cite{wang2022broadband} and \cite{lv2023broadband} designed $\mathrm{VO_{2}}$-based RISs with ultra-wideband beam steering capability, operating in the frequency bands of 0.218–0.232 THz and 0.735–0.965 THz, respectively, with steering angles exceeding $\mathrm{50^\circ}$. Although $\mathrm{VO_{2}}$ exhibits excellent phase-change properties, its application is hindered by the requirement for heating. In \cite{wu2020liquid}, a liquid crystal-based RIS was developed, working at 683 GHz and supporting beam steering within a range of $32^\circ$. Furthermore, the authors in \cite{li2023modulo} introduced a liquid crystal-based RIS capable of two-dimensional beam steering for incident waves at the 0.728 THz band. However, highly directional THz beams and the unexpectedly long switching time of tunable materials make it difficult for RISs in these studies to serve multiple users simultaneously. To address this issue, the authors in \cite{chen2024liquid} designed a liquid crystal-based 0.29 THz RIS, which could simultaneously redirect incident beams to multiple desired directions, thereby enabling service to multi-users concurrently.

Most of the aforementioned studies focus solely on the design of THz band RIS,  while research on the joint design of RIS and THz communication systems is rare. In these studies, the THz source signal is merely a single-tone signal, which does not carry any transmitted information. The performance of the designed RIS in practical THz communication systems requires further experimental validation. Additionally, in multi-beam RIS-assisted communication systems, the reflected sub-beams carry identical information. A key challenge remains: how to leverage these sub-beams to deliver distinct information to multiple users simultaneously?

\subsection{Our Contributions}
To cope with the above challenges, we designed a THz band RIS-assisted communication prototype system. 
A corresponding multi-user baseband transmission scheme was proposed, in which the modulated signal was further up-converted to the 220 GHz band for transmission.
Our contributions can be summarized as follows:

%\begin{enumerate}
\begin{itemize}
	\item{A liquid crystal-based 220 GHz RIS is designed. Unlike traditional multi-bit adjustment schemes, it possesses the capability to deflect incident waves continuously over a range of $110^\circ$. Besides, it can not only redirect the incident beam to a single desired angle but also split the incident beam into multiple sub-beams, steering the highly directional terahertz waves to several different desired directions, thereby supporting simultaneous communication for multiple users.}
	\item{A multi-user-enabled THz transmission scheme is implemented. To address the issue that the sub-beams reflected by the RIS carry the same information, making it challenging to serve multiple users simultaneously, we allocate separate OFDM resources to different users. In this mode, the transmitter sends an OFDM multiplexing data frame, and different users can receive their respective data independently and without interference.}
	\item{A beamforming method based on the stochastic parallel gradient descent (SPGD) algorithm is developed, which is applicable to both single-beam and multi-beam scenarios. Furthermore, we design a power feedback adjustment method for multi-beam applications to address the problem of low received power for sub-beams in certain directions.}
	\item{A prototype of RIS-assisted THz communication system equipped with a power feedback link is built. Experimental results demonstrate that our system could effectively support simultaneous communication for multiple users, with an effective data rate of up to 2.341 Gbps. To the best of our knowledge, this is the first prototype that integrates RIS into the practical THz band and realizes multiple-user over-the-air image transmission.}
\end{itemize}

\subsection{Organization}
The organization of the rest of this paper is as follows. In Section \ref{II}, we describe the design details of the 220 GHz THz-band RIS. In Section \ref{III}, we introduce the system architecture and multi-user transmission strategy of the designed RIS-assisted THz communication system. In Section \ref{IV}, we discuss the beamforming method and power feedback adjustment for the RIS. In Section \ref{V}, we validate the performance of the designed system through experiments conducted in various scenarios. Finally, in Section \ref{VI}, we conclude our work.

\begin{figure}
	\centering
	\includegraphics[width=0.30\textwidth]{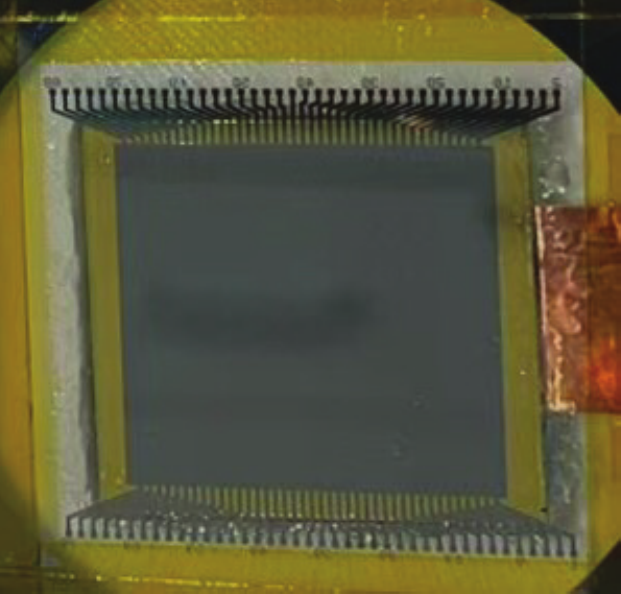}\vspace{0em}\\
	\caption{Appearance of the designed THz band RIS with 80 elements.}\label{fig1}\vspace{0.0em}
\end{figure}

\section{Design of the 220 GHz RIS }\label{II}
In this section, we elaborate on the structure, frequency characteristics, and control methods of the 220 GHz RIS.

\subsection{Design of 220 GHz RIS}
\begin{figure}
	\centering
	\includegraphics[width=0.42\textwidth]{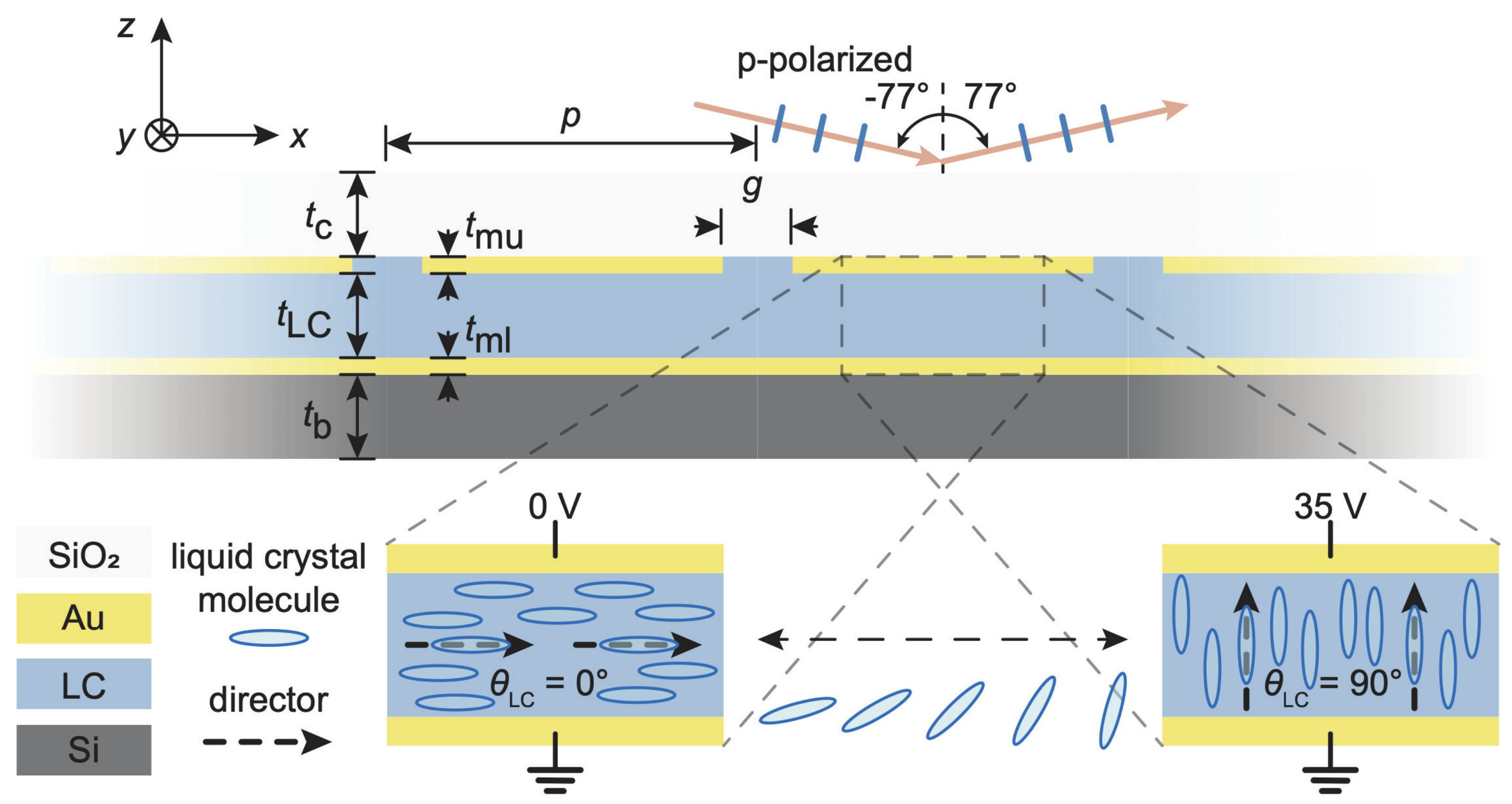}\vspace{0em}\\
	\caption{Cross-sectional structure of reflective elements of RIS.}\label{fig2}\vspace{0.0em}
\end{figure}

We designed a 220 GHz band RIS based on liquid crystal materials, composing 80 individual units. Fig. \ref{fig1} shows the appearance of the assembled RIS. Fig. \ref{fig2} illustrates the cross-sectional view of reflective elements, which consists of five sub-layers. The bottom layer is a silicon substrate, providing stable mechanical support; the fourth and the second layer are both thin film layers of gold; the third layer is a dielectric layer of liquid crystal, and the top layer is a silica cover plate that protects the device. The two gold-plated metal film layers act as two electrodes. When the electric field between the two electrodes changes, the orientation of liquid crystal molecules, as $\theta_{LC}$ shown in Fig.\ref{fig2}, will rotate and modify the dielectric constant and impedance of the liquid crystal layer. Consequently, the reflection coefficient of the reflective element will vary, by which the amplitude and phase of the incoming signal are adjusted. In order to design RIS operating at 220 GHz, several parameters, such as the period of the gold-plated pattern and thickness of the liquid crystal layer, need to be appropriately set. Some critical parameters are listed in Tab. \ref{tab1}.

\subsection{Frequency Response of the Designed RIS}
To verify the performance of the design, we simulated the phase responses of a single RIS element with liquid crystal molecules in different orientations, shown in Fig.\ref{fig3}. We observed that at a frequency of 220 GHz, tuning liquid crystal molecules produced a phase response ranging from 0 to $\mathrm{255^\circ}$. Notably, frequency sweep curves indicate a very wide linear range for the phase response. It can be considered as having a flat frequency response within the 6 GHz bandwidth centered around 220 GHz. This indicates that our RIS can effectively control wideband signals. 

\subsection{Control Method of the Designed RIS}
\begin{figure}
	\centering
	\includegraphics[width=0.42\textwidth]{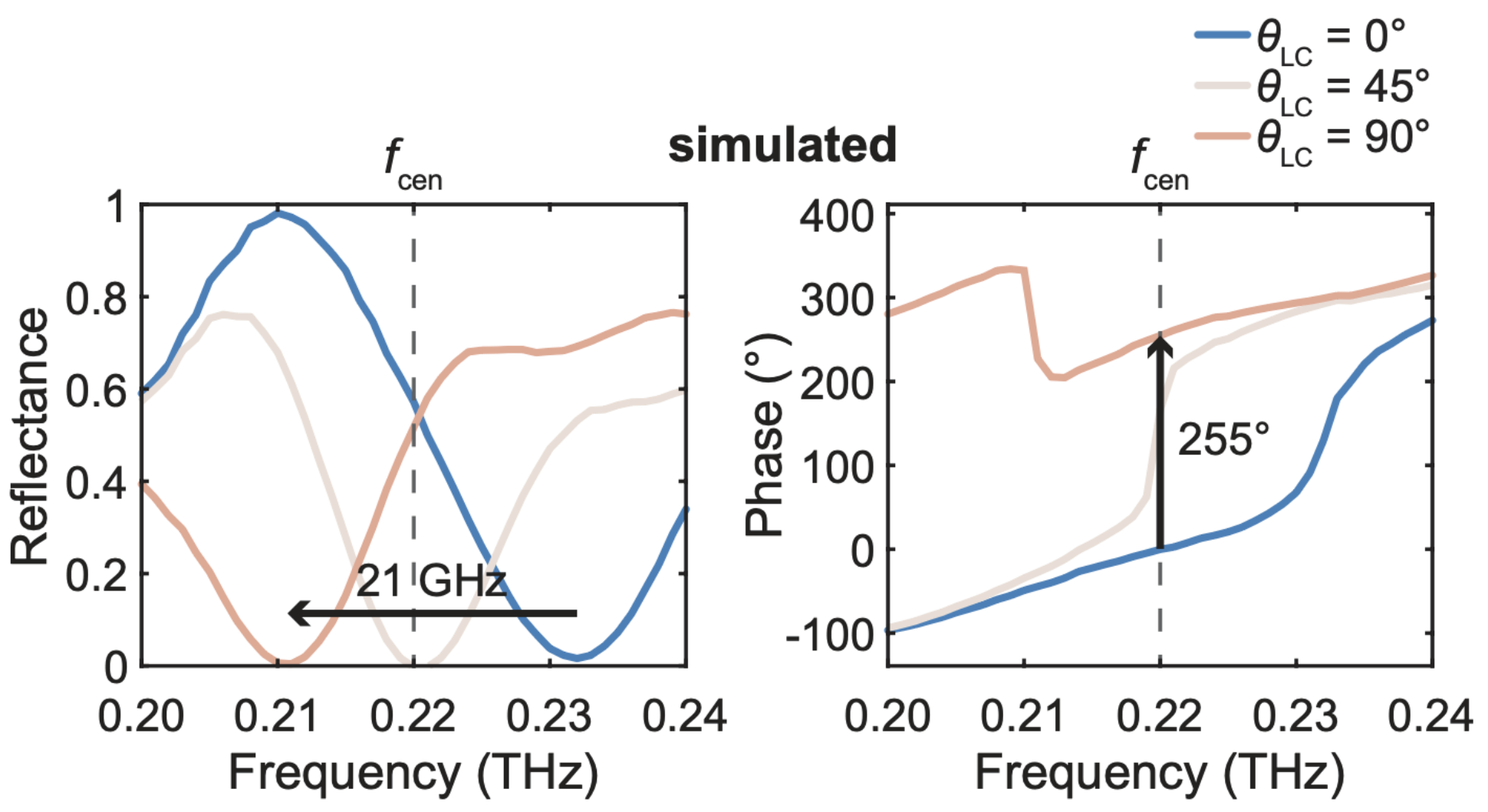}\vspace{0em}\\
	\caption{Simulated amplitude-frequency and phase-frequency response with different orientations of liquid crystal molecules.}\label{fig3}\vspace{1em}
	\centering
	\includegraphics[width=0.42\textwidth]{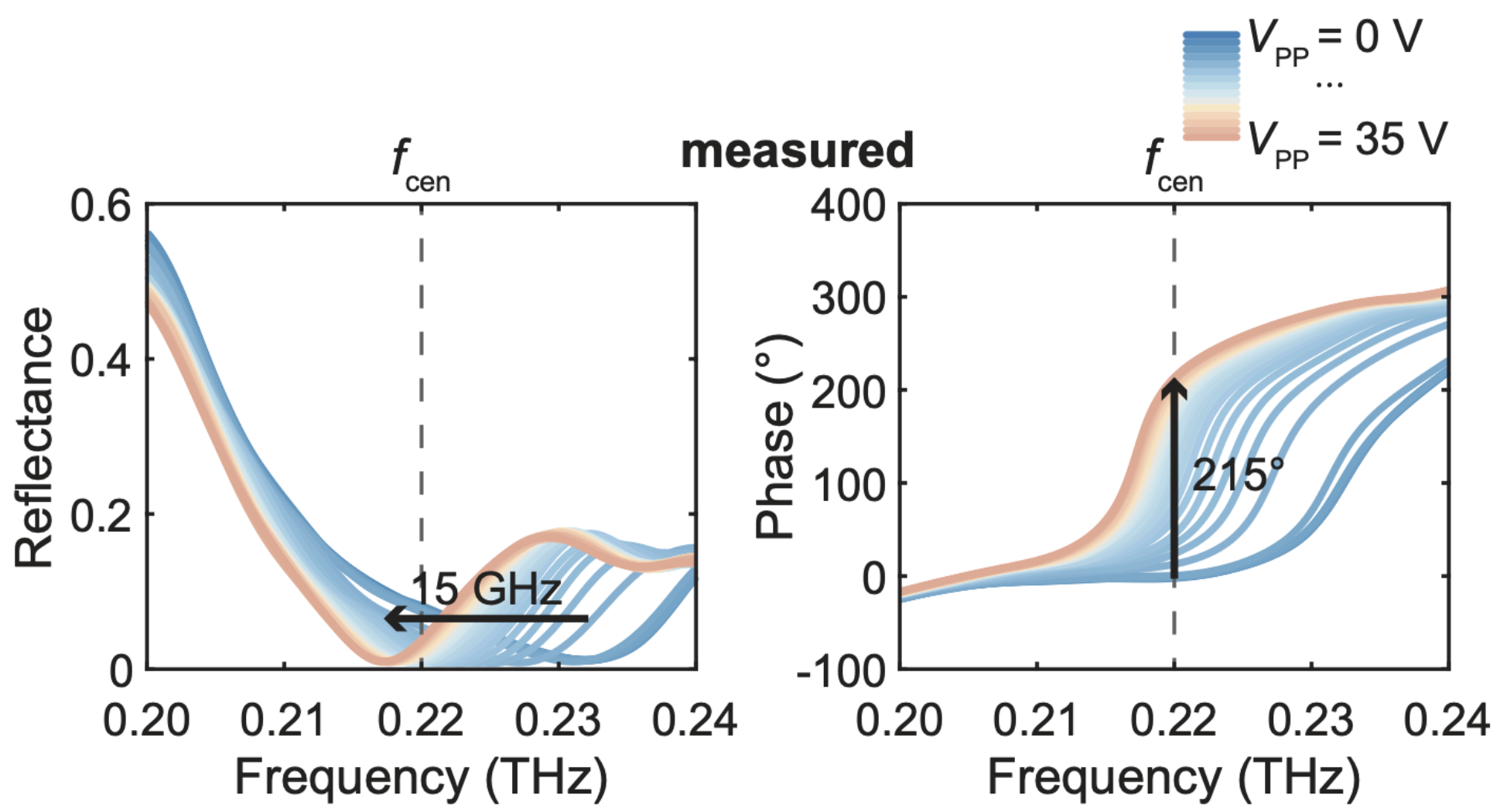}\vspace{0em}\\
	\caption{Measured amplitude-frequency and phase-frequency response with different bias voltages applied to the RIS.}\label{fig4}\vspace{0.0em}
\end{figure}

The orientation of liquid crystal molecules can be adjusted by applying different bias voltages. In our system, liquid crystal molecules can be tilted from 0 to $\mathrm{90^\circ}$ by continuously varying the voltage from 0 to 35V. Fig.\ref{fig4} illustrates the amplitude and phase responses of a single reflective element under different bias voltages, which match the simulation results. In our system, an FPGA is employed to control the bias voltages of all reflective elements simultaneously. The demanded bias voltage for each element are pre-calculated at the transmitter based on the desired reflection coefficients and then sent to the FPGA\footnote{In general, the transmission of control information can be carried out using wired or wireless methods. For simplicity, we employed wired transmission in this work.}. The FPGA is powered by a 35V DC supply and could assign different bias voltages ranging from 0V to 35V to each element according to the received commands.

\begin{table}[t]
	\footnotesize
	%	\vspace{-0.5cm}
	\caption{Parameters of the designed RIS }\vspace{-1em}
	\begin{center}
		\begin{tabular}{ll}
			\Xhline{0.7px}
			Parameters & Values \\
			\hline
			The thickness of the upper gold layer, $t_{mu}$ &$0.2\ \mu m$ \\
			The thickness of the lower gold layer, $t_{ml}$& $0.1\ \mu m$\\
			The thickness of liquid crystal layer, $t_{LC}$ &$16\ \mu m$ \\
			The thickness of silicon substrate, $t_b$ &$500\ \mu m$ \\
                The thickness of silica cover plate, $t_c$ & $1000\ \mu m$ \\
			The gap width between two adjacent patterns, $g$ & $7\ \mu m$\\
			The period of a single element, $p$ & $317\ \mu m$\\
			\Xhline{0.7px}
		\end{tabular}
	\end{center}\label{tab1}\vspace{0em}
\end{table}

\section{Design of RIS Assisted Terahertz Communication Systems }\label{III}
In this section, we describe the architecture of the designed RIS-THz communication system and elaborate on the functionality of each component. Specifically, we introduce a multi-user-oriented OFDM resource allocation scheme.

\subsection{The Framework of the System}
\begin{figure*}
	\centering
	\includegraphics[width=0.90\textwidth]{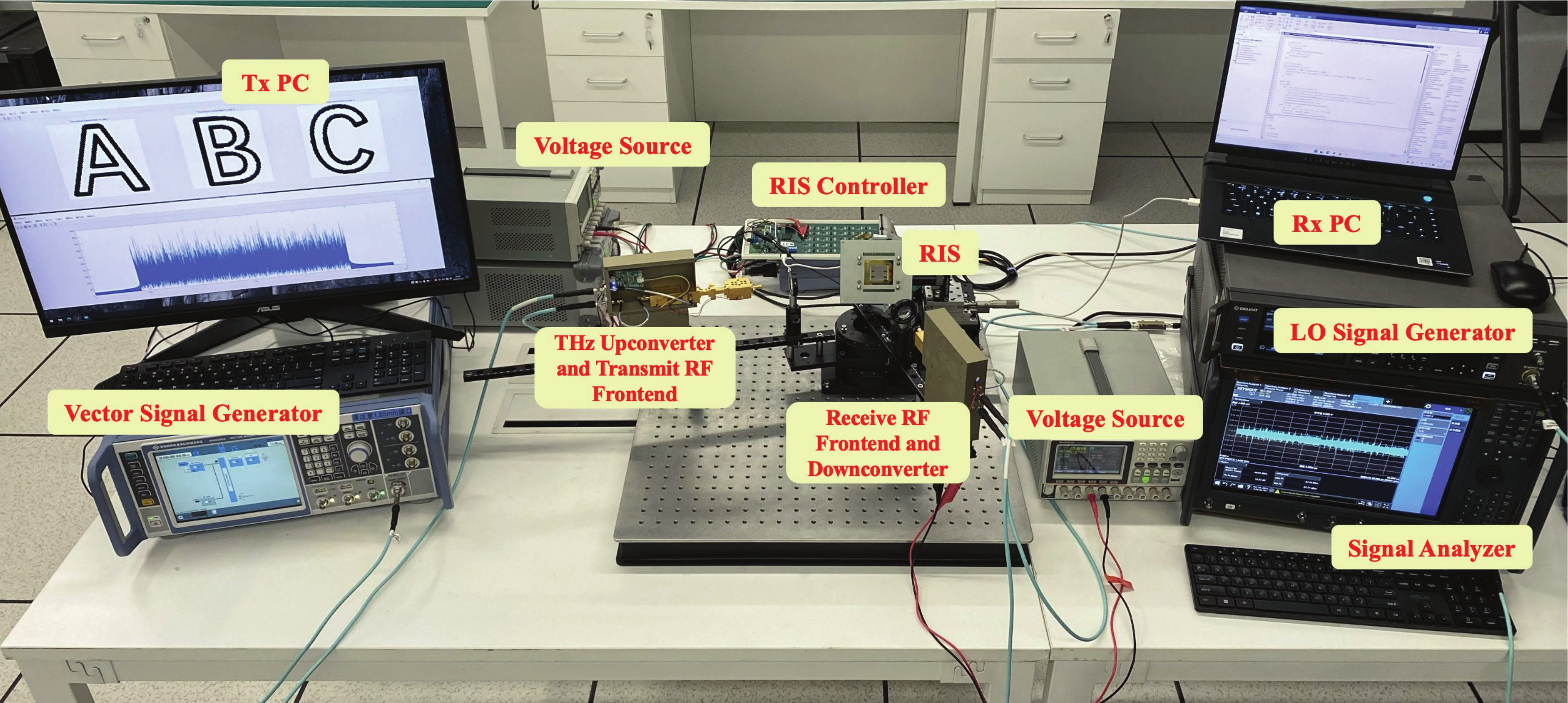}\vspace{0em}\\
	\caption{Prototype of the designed RIS-aided 220 GHz communication system, which consists of five parts, baseband-to-IF transmitter, THz up-converter and transmit RF frontend, RIS and its controller, receive RF frontend and THz down-converter, and IF-to-baseband receiver, respectively. }\label{fig5}\vspace{0.0em}
\end{figure*}

\begin{figure*}
	\centering
	\includegraphics[width=0.90\textwidth]{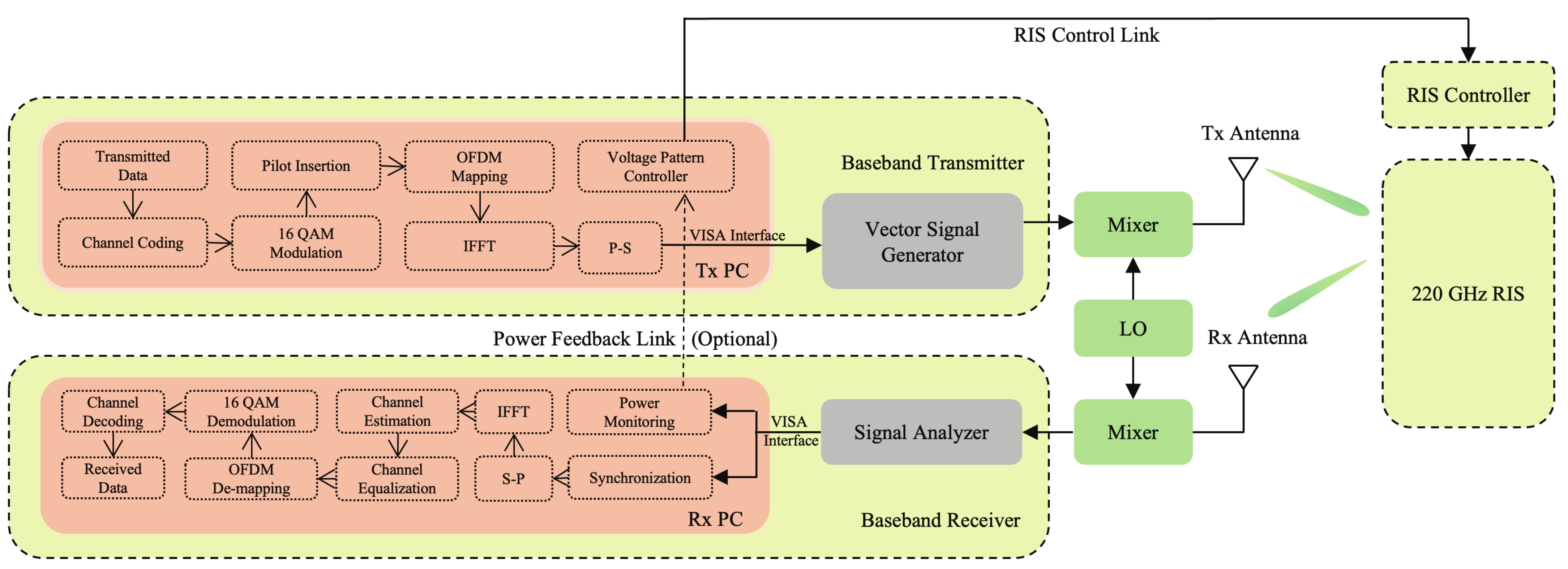}\vspace{0em}\\
	\caption{Architecture of the designed RIS-aided 220 GHz communication system.}\label{fig6}\vspace{0.0em}
\end{figure*}

The designed RIS-THz communication system prototype is shown in Fig. \ref{fig5}, and the architecture of the system is illustrated in Fig. \ref{fig6}. The system primarily consists of five parts: the baseband-to-intermediate frequency (IF) transmitter, the THz upconverter and transmit RF frontend, the RIS and its controller, the receive RF frontend and THz downconverter, and the IF-to-baseband receiver. We elaborate on each part as follows. Some specifications of the hardware components used in our system are provided in Tab. \ref{tab2}.

\textbf{Baseband-to-IF Transmitter:} This includes a host personal computer (PC) and a vector signal generator (VSG), connected via an Ethernet interface and using the virtual instrument software architecture (VISA) protocol for data transmission. As shown in Fig.\ref{fig6}, the host PC is responsible for generating data bits and performing channel encoding, modulation, pilot insertion, OFDM resource mapping, IFFT, and parallel-to-serial conversion. The baseband data generated by the PC is then sent to the VSG, which converts it into an IF RF signal and outputs it to the terahertz upconverter.

\textbf{THz Up-Converter and Transmit RF Frontend:} This is composed of a cascaded up-converter, a band-pass filter, a power amplifier, and a horn antenna. The cascaded up-converter mixes the IF signal with the local oscillator (LO) signal to generate a double-sideband signal around 220 GHz, with one sideband filtered out to leave a single band signal. The amplified signal is radiated into free space through the horn antenna.

\textbf{RIS and its Controller:} The designed liquid crystal metasurface serves as the RIS to deflect the highly directional terahertz beam to one or multiple specified directions. The controller is implemented by an FPGA, which receives control signals from the transmitter and configures the bias voltage for each RIS element accordingly.

\textbf{Receive RF Frontend and THz Down-Converter:} This mainly includes a horn antenna, a low-noise amplifier (LNA), a down-converter, and a band-pass filter. The signal received by the antenna is amplified and down-converted to an IF signal\footnote{The local oscillator signals used at transmitter and receiver need to be strictly in the same frequency and phase. For simplicity, in this work, we use the same source on both the transmitter and receiver.}, with the unexpected sideband filtered out before being sent to the IF-to-baseband receiver.

\begin{table}[t]
	\footnotesize
	%	\vspace{-0.5cm}
	\caption{Specifications of the Hardware used in our work}\vspace{-1em}
	\begin{center}
		\begin{tabular}{ll}
			\Xhline{0.7px}
			Instruments & Specifications \\
			\hline
			Vector Signal Generation & Rohde-Schwarz SMW200A\\
			Signal Analyzer & Keysight N9042B\\
			LO for Frequency Multiplier & 12.5 GHz \\
                Transmitted Filter Pass Band & 205 GHz to 225 GHz\\
			CPU of host PC & Intel i7-1250K, Intel i9-12900H \\
			Beamwidth of Horn Antenna & $7^\circ$ \\
                Transmitted Power Amplifier Gain & $18$ dB \\
                Received Power Amplifier Gain & $27$ dB\\   
			\Xhline{0.7px}
		\end{tabular}
	\end{center}\label{tab2}\vspace{-1.5em}
\end{table}

\textbf{IF-to-Baseband Receiver:} This consists of a signal analyzer (SA) and a receiving host PC. The SA converts the IF signal to the baseband and then captures its I-Q waveform and calculates its power. The I-Q waveform is sent to the signal processing module for synchronization, serial-to-parallel conversion, and other steps until the transmitted data bits are demodulated. The detected signal power is further sent to the power monitoring module and fed back to the transmitter, which helps the transmitter to select a more suitable voltage pattern\footnote{This feature is optional. In a multi-user scenario, if the quality of signals received by different users varies significantly, the feedback link will be activated to allow for more precise adjustments. Conversely, if the signal quality across multiple users has reached the desired level, no adjustment via the feedback link is necessary.}.

\subsection{Transmission Strategy for Multi-Users}
In this section, we design a multi-user-oriented OFDM transmission scheme. The THz band offers abundant frequency resources, and with the high transmission bandwidth supported by the designed RIS, the OFDM frame used in our system could contain a larger number of time-frequency resource blocks \cite{shafie2021spectrum}. In our prototype scheme, each OFDM data frame is defined as 120 time-frequency resource blocks, each of which contains 14 OFDM symbols and 12 subcarriers. When there is only a single user in the system, it can occupy all the resources for data transmission, while multiple users are present, the whole resource blocks can be divided into several parts and allocated to users in different directions. In our demonstration, we consider a scenario where the incident beam is split into three directions. The allocation of OFDM resources is shown in Fig. \ref{fig7}. The yellow resource units serve as pilot symbols, and the green/pink/blue resource units are allocated for data transmission to users in three different directions, respectively. Users in each direction only need to receive and demodulate the data in their allocated resource units, ensuring interference-free simultaneous communication across multiple directions.

\section{Beamforming Method and Feedback Control}\label{IV}
In this section, we presented a RIS-aided beamforming strategy. Moreover, to achieve fine-tuning of the sub-beams, we designed a power feedback adjustment method.

\subsection{RIS Beamforming Design}
Beamforming design is a crucial issue in RIS-assisted communication systems. By appropriately configuring the RIS's phase shift matrix, the incident beam can be steered in the desired direction. In most existing studies, an optimization problem is typically formulated, and the theoretical optimal phase shift can be obtained by solving the formulated problem \cite{yan2023beamforming,10184278,10684280}. However, these methods often rely on the known RIS-related channel state information (CSI), which is very challenging to acquire in practical communication systems.

\begin{figure}
	\centering
	\includegraphics[width=0.45\textwidth]{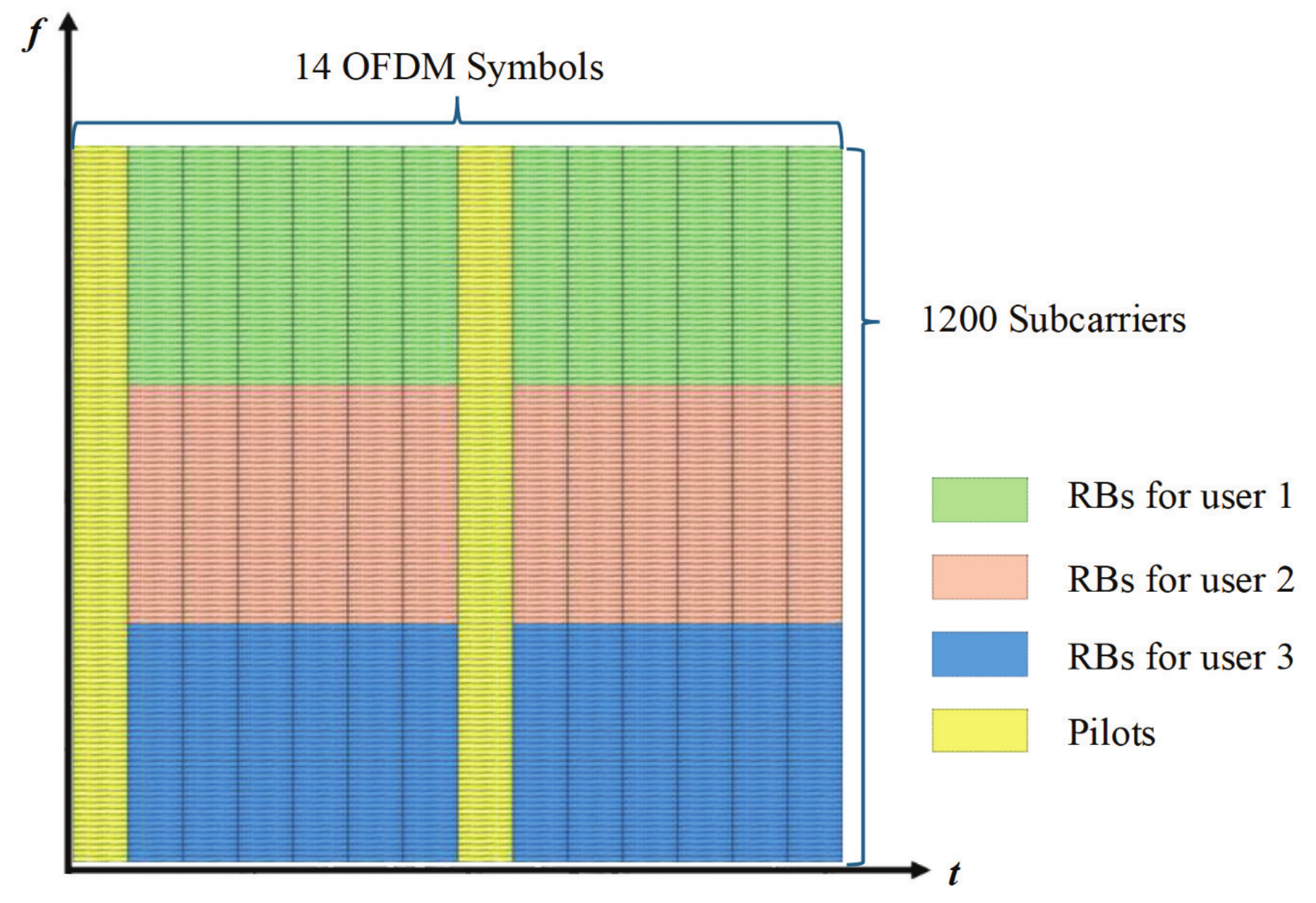}\vspace{0em}\\
	\caption{OFDM resource allocation scheme employed in our work.}\label{fig7}\vspace{0.0em}
\end{figure}

Furthermore, in practical RIS-assisted communication systems, the RIS phase shift needs to be mapped to the controlling commands (typically the amplitude of control voltages, duty cycle, etc.). The phase shift obtained through the aforementioned methods requires further calculation before it can be applied practically. Moreover, in real-world systems, due to the noise and nonlinearity inside the devices and the correlation in the electric field distribution between neighboring elements, the optimal RIS phase shift obtained by the optimization theory is often not truly optimal, especially in high-frequency bands.

To address these problems, in this paper, we adopt a stochastic parallel gradient descent (SPGD) optimization algorithm, which has been widely used in engineering \cite{vorontsov1998stochastic}. Unlike traditional optimization theory-based methods, this approach could directly obtain the optimal control voltage pattern, which makes it highly practical. Specifically, when a beam reflected to the ${\theta_d}^\circ$ direction is expected,  the initial phase shift $\boldsymbol{\varphi}_0$ can be computed using the grating equation 
\begin{align}
     \boldsymbol{\varphi}_0(i) = \mathit{i}\times\mathit{k}\times\mathit{p}\times\left(\sin \theta_{in} +\sin \theta_d \right) , i=1,2,\cdots,N,
\end{align}
where $\mathit{i}$ is the index of reflective elements in RIS, $\mathit{k}$ is the wavenumber at a certain frequency, $p$ is the grating constant, ${\theta_{in}}$ is the direction of the incoming signal, and $N$ is the number of elements of RIS.

Based on this initial phase shift, we can obtain the initial voltage pattern $\mathbf{v}_0$, which is a vector with the same length as $\boldsymbol{\varphi}_0$ and contains the bias voltage of each element. As previously mentioned, due to the device characteristics and coupling effects between neighboring elements, this initial voltage pattern is often suboptimal and requires further optimization in the practical systems. In this process, the realistic received power (RRP) in the desired direction serves as the evaluation metric. When other conditions remain fixed, RRP is only a function of the control voltage pattern, which can be expressed as
\begin{align}
	\mathit{y} = \mathit{f}\left( \mathbf{v}\right), 
\end{align}
where $\mathit{y}$ represents the RRP with the voltage pattern $\mathbf{v}$. In fact, the process of obtaining the optimal voltage pattern for the desired direction could be viewed as a beamforming process. First, we measure the RRP under the initial voltage pattern $\mathbf{v}_0$ and regard it as $\mathit{y}_0$. Then, a small random perturbation $\Delta \mathbf{v}$ is applied to the initial voltage pattern $\mathbf{v}_0$ and obtain the corresponding RRP as $\mathit{f}(\mathbf{v}_0+\Delta\mathbf{v}) $. Next, apply the opposite perturbation $-\Delta\mathbf{v}$ to the initial voltage pattern $\mathbf{v}_0$ and obtain the corresponding RRP as $\mathit{f}(\mathbf{v}_0-\Delta\mathbf{v} ) $. Consequently, the updated voltage pattern can be calculated as $	\mathbf{v}_{new} = \mathbf{v}_0 + \mathit{g} \Delta\mathbf{v} \left[\  \mathit{f}(\mathbf{v}_0+\Delta\mathbf{v})-\mathit{f}(\mathbf{v}_0-\Delta\mathbf{v} )\ \right]$, where $\mathit{g}$ is a scaling constant. A more general form for iterations can be expressed as
\begin{align}\label{vnew}
\mathbf{v}_{t+1} = \mathbf{v}_t + \mathit{g} \Delta\mathbf{v}_t \left[\  \mathit{f}(\mathbf{v}_t+\Delta\mathbf{v}_t)-\mathit{f}(\mathbf{v}_t-\Delta\mathbf{v}_t )\ \right],
\end{align}
where $t$ represents the iteration number. When $t$ reaches a preset threshold or no further improvement on RRP can be made, the optimal voltage pattern is obtained. It is worth noting that this process is conducted offline. Once the beamforming process is completed, the optimal voltage pattern for a particular direction can be stored and applied directly when in need. 

\begin{algorithm}[t]
	\small
	\caption{\small{SPGD based beamforming method}}
	\label{algo1}
	\textbf{Input:} The pre-calculated initial voltage pattern $\mathbf{v}_0$, the set of all directions $K$, the scale constant $g$, and the number of iterations $J$. 
	\begin{algorithmic}[1]
		\State \text{Initialize the iteration index $t=0$ and initialize $\mathbf{v}_t=\mathbf{v}_0$ ;}
		\Repeat
		\State Generate a random perturbation patten $\Delta \mathbf{v}_t $;
		\State Measure $\mathit{f}(\mathbf{v}_t+\Delta\mathbf{v}_t) or \sum_{k\in K}\mathit{f}_k(\mathbf{v}_t+\Delta\mathbf{v}_t)$;
		\State Measure $\mathit{f}(\mathbf{v}_t-\Delta\mathbf{v}_t) or \sum_{k\in K}\mathit{f}_k(\mathbf{v}_t+\Delta\mathbf{v}_t) $ ;
		\State Update the voltage pattern by (\ref{vnew}) or (\ref{vnewm});
		\State Update $t \leftarrow t+1$; 
		\Until $t=J$ ;
	\end{algorithmic}
	\textbf{Output:} The optimal optimal voltage pattern $\mathbf{v}^\star$.
\end{algorithm}

On the other hand, when the incident beam needs to be steered toward multiple directions, the summed average voltage pattern on all desired directions can be used as an initial pattern. Specifically, instead of using the RRP on a single desired direction as the evaluation metric, the sum of RRP on all desired directions is used, which can be expressed as
\begin{align}\label{multi-train}
	\mathit{y} = \sum_{k\in K}\mathit{f}_k\left( \mathbf{v}\right), 
\end{align}
where $k$ is the index of the sub-directions and $K$ is the set of all directions. In this case, (\ref{vnew}) can be corrected as 
\small
\begin{align}\label{vnewm}
	\mathbf{v}_{t+1} = \mathbf{v}_t + \mathit{g} \Delta\mathbf{v}_t \left[\ \sum_{k\in K} \mathit{f}_k(\mathbf{v}_t+\Delta\mathbf{v}_t)-\sum_{k\in K}\mathit{f}_k(\mathbf{v}_t-\Delta\mathbf{v}_t )\ \right].
\end{align}

\normalsize
The rest of the beamforming method is identical to the single-beam case. Algorithm \ref{algo1} summarizes the complete beamforming method used in our work. It is worth noting that employing (\ref{multi-train}) as the evaluation metric for multi-beam cases may encounter certain problems, on which we will elaborate in the next section.

\subsection{Feedback Control of Received Power}
\begin{algorithm}[t]
	\small
	\caption{\small{RRP feedback adjustment method}}
	\label{algo2}
	\textbf{Input:} The measured RRP on the all sub-directions $\mathbf{p}$, and the current weights of all sub-directions $\mathbf{w}=[w_1,\cdots w_k]$ . 
	\begin{algorithmic}[1]
		\State Calculate the minimum component of $\mathbf{p}$ and acquire its index $i$;
		\State Update $\mathbf{w}(i) = \mathbf{w}(i)+1$;
		\State Update $\mathbf{v}$ with the updated $\mathbf{w}(i)$ using Algorithm \ref{algo1};
	\end{algorithmic}
	\textbf{Output:} The feedback voltage pattern $\mathbf{v}_{fe}$.
\end{algorithm}

Employing the equation (\ref{multi-train}) as the evaluation metric may present a problem in that although the sum of RRP on all desired directions converges to the maximum, the RRP on a certain sub-direction may not achieve its maximum. The worst case is that the RRP on different sub-directions varies significantly. Furthermore, in practical communication systems, the channels in different directions are not identical and may differ considerably. The voltage patterns obtained through the previous method may not fully adapt to these variations. Therefore, it is necessary to devise a method to adjust the RRP among multiple sub-beams more precisely. 

In our work, we introduce a multi-direction beam weight adjustment method based on power feedback. Firstly, taking the weights of the sub-beams in each direction into account, the previous metric (\ref{multi-train}) can be corrected as 
\begin{align}\label{weight}
	\mathit{y} = \sum_{k\in K}\mathit{w}_k \mathit{f}_k\left( \mathbf{v}\right), 
\end{align}
where $\mathit{w}_k$ represents the weight on each direction. It is evident that the voltage pattern obtained by Algorithm \ref{algo1} is essentially the case when the weights of all directions are the same. In our system, we use the voltage pattern obtained by Algorithm \ref{algo1} as the initial voltage pattern. Then the user in each direction feeds back its RRP to the transmitter. The transmitter selects the direction with the minimum RRP and sends a voltage pattern with greater weight for that direction to the RIS controller, thus completing one feedback procedure. Algorithm \ref{algo2} summarizes the RRP feedback adjustment method.

\subsection{Beamforming Setup}
\begin{figure}
	\centering
	\includegraphics[width=0.40\textwidth]{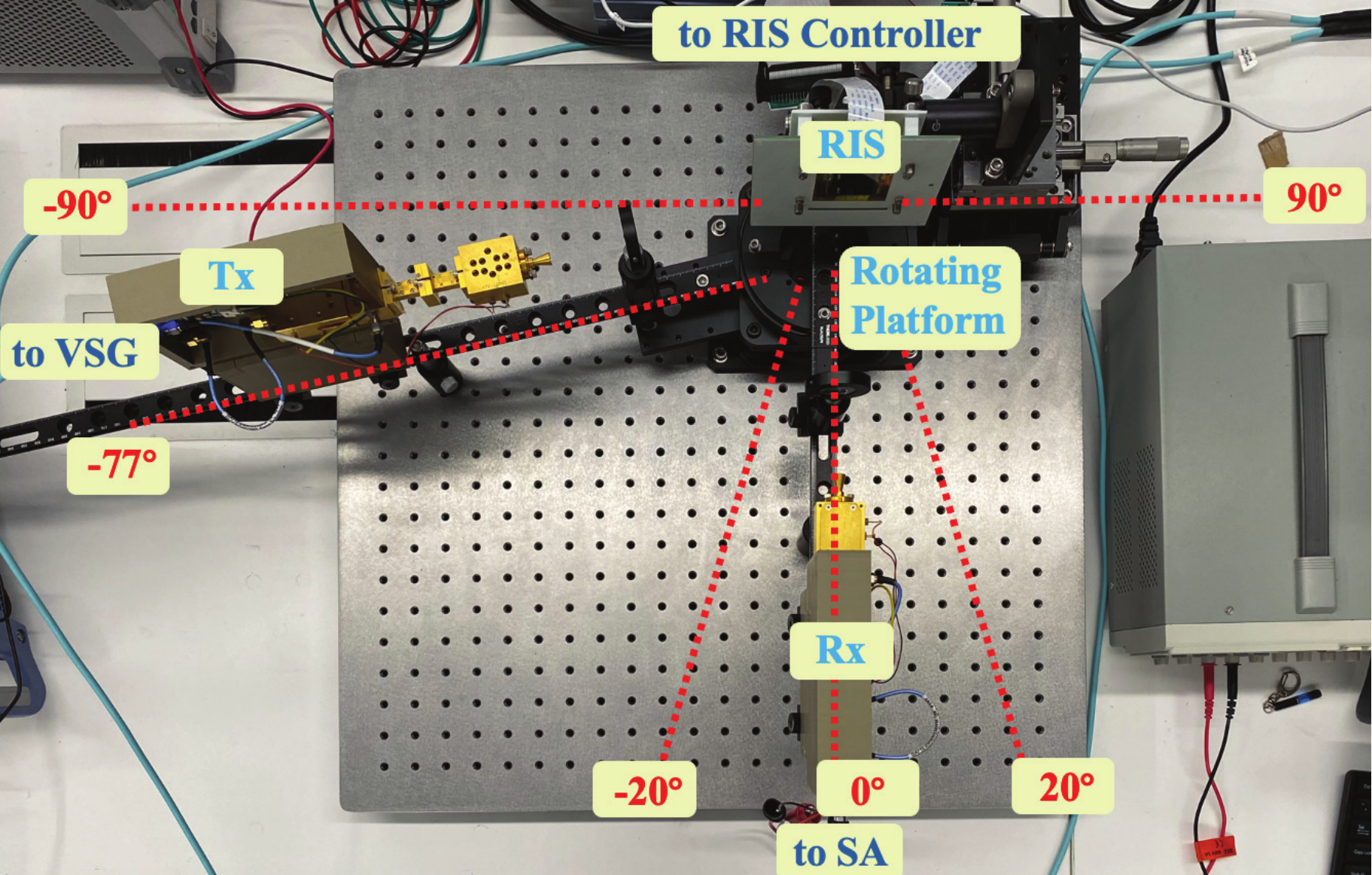}\vspace{0em}\\
	\caption{Beamforming scenario and experimental environment in our work. }\label{fig8}\vspace{0.0em}
\end{figure}

The beamforming scenario is shown in Fig. \ref{fig8}. The designed THz band RIS is fixed on the central axis of an electronically controlled rotating platform and connected to the RIS controller. The transmitter and receiver are fixed on two rotating rails. Theoretically, the closer the transmitter is to $-90^\circ$, the wider the transmission range of the signal after RIS reflection. However, when the transmitter is positioned very close to $-90^\circ$, the transmitted electromagnetic wave travels almost parallel to the RIS, such that it cannot be effectively reflected by the RIS. To avoid this, after our experiments, $-77^\circ$ is a very appropriate direction to deploy the transmitter, while the receiver can rotate freely between $-50^\circ$ to $90^\circ$. In the beamforming process, a single tune at 220 GHz is transmitted, and the RRP of the received signal is measured at the receiver. Then it follows the procedure outlined in Algorithm \ref{algo1}. In the single-beam-forming process, we constructed 12 beams in the directions from $-50^\circ$ to $60^\circ$ with a $10^\circ$ interval. In the multi-beam-forming process, we constructed beams on four groups of directions including $(-50^\circ, -20^\circ, 50^\circ)$, $(40^\circ, 50^\circ, 60^\circ)$, $(-20^\circ, 0^\circ, -20^\circ)$, and $(0^\circ, 10^\circ, 30^\circ)$.

\section{Experimental Results}\label{V}
In this section, we first introduce the experimental details of beamforming. Then we evaluate the performance of the designed RIS-THz communication system under both single-user and multi-user scenarios. Additionally, through the transmission of a set of images, we vividly demonstrate that the RRP feedback method could precisely adjust the quality of received signals in certain directions.

\subsection{Verification of Single User Scenario}
\begin{figure}
	\centering
	\includegraphics[width=0.43\textwidth]{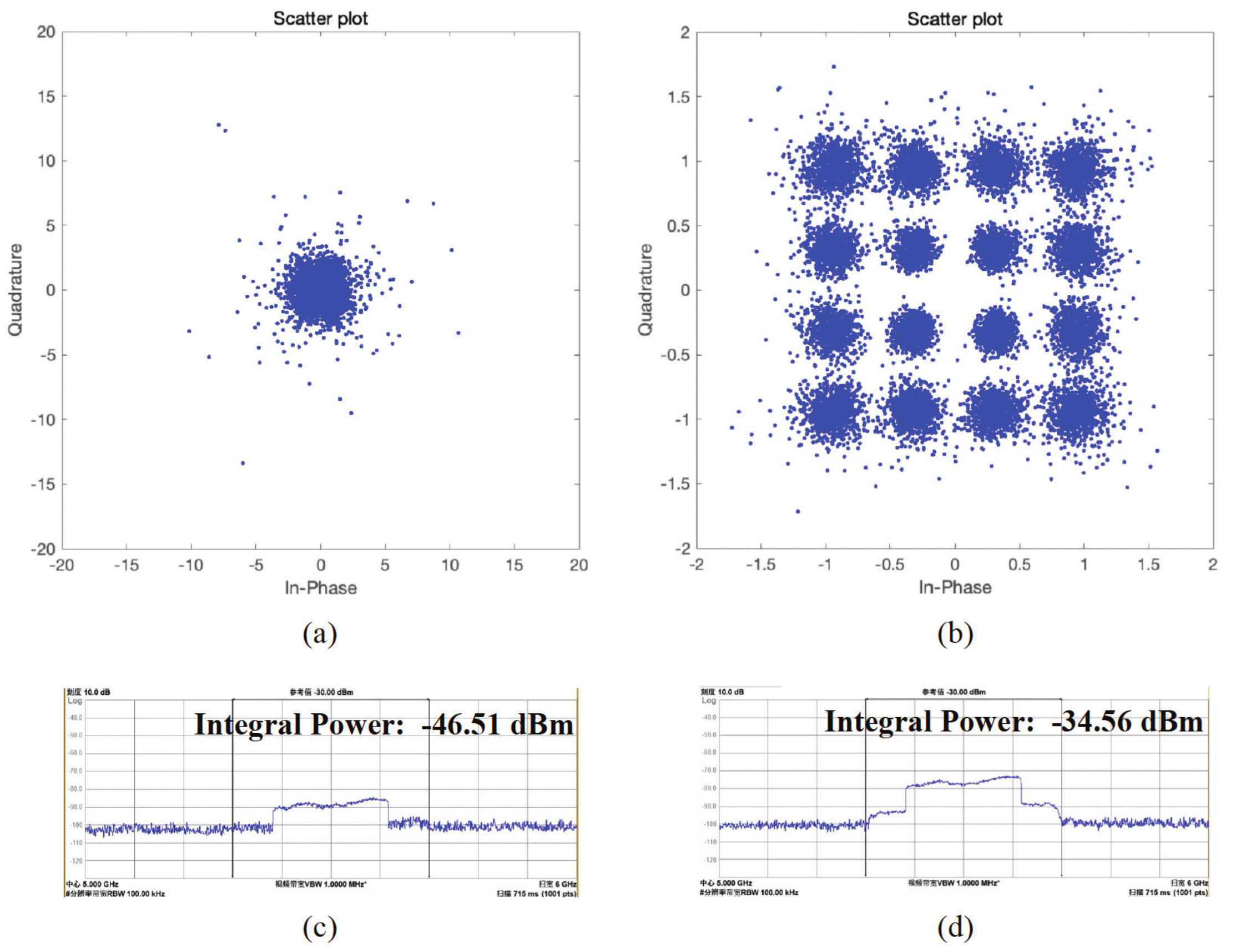}\vspace{0em}\\
	\caption{Constellation diagram and spectrum on the $60^\circ$, (a) constellation diagram without the bias voltage; (b) constellation diagram with the bias voltage; (c) spectrum without the bias voltage; (d) spectrum with the bias voltage.}\label{fig10}\vspace{0.0em}
\end{figure}

\begin{figure}
	\centering
	\includegraphics[width=0.43\textwidth]{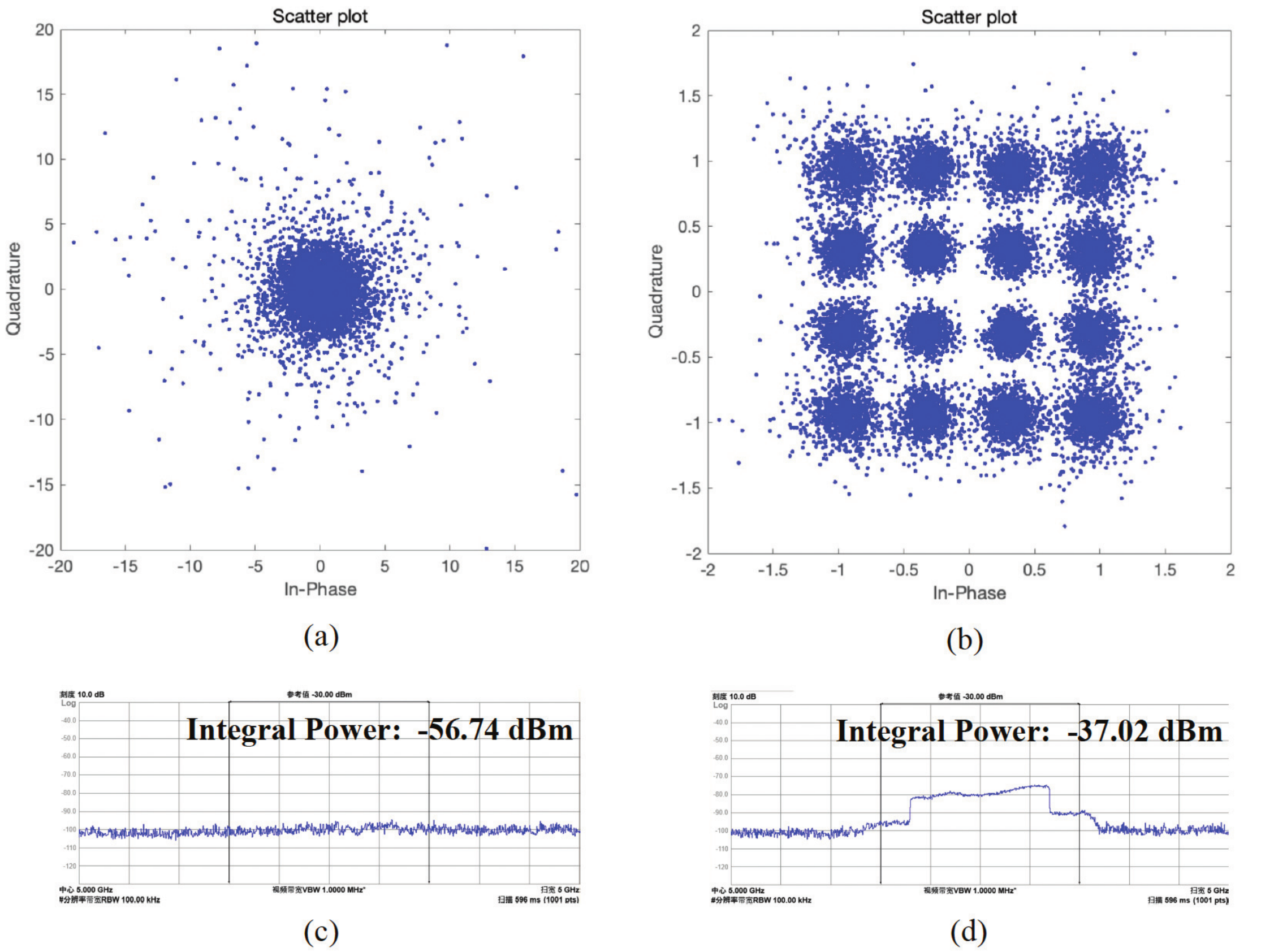}\vspace{0em}\\
	\caption{Constellation diagram and spectrum on the $-20^\circ$, (a) constellation diagram without the bias voltage; (b) constellation diagram with the bias voltage; (c) spectrum without the bias voltage; (d) spectrum with the bias voltage.}\label{fig11}\vspace{0.0em}
\end{figure}

In this subsection, we validate the performance of the designed system in a single-user scenario. In this scenario, the RIS steers the incident signal in a single direction. The physical layer parameters adopted for transmission are presented in Tab.\ref{tab3}. The experimental environment is the same as depicted in Fig.\ref{fig8}. With this configuration, we sequentially measured the RRP on the directions ranging from $-50^\circ$ to $60^\circ$ with a $10^\circ$ interval. The bit error rate (BER) is calculated at the receiver. We select $60^\circ$ and $-20^\circ$ as two representative directions, and their spectra and constellation diagrams are shown in Fig.\ref{fig10} and Fig.\ref{fig11}, respectively.

Fig.\ref{fig10}(a) shows the constellation diagram of the received signal at $60^\circ$ without applying the bias voltage to the RIS, while Fig.\ref{fig10}(b) presents the constellation diagram when the bias voltage is applied. Obviously, the constellation diagram becomes significantly clearer with the application of bias voltage. Fig.\ref{fig10}(c) displays the spectrum of the received signal without the bias voltage, and Fig.\ref{fig10}(d) shows the spectrum with the bias voltage applied. The results indicate that the introduction of RIS can effectively redirect the signal in a desired direction and increase the RRP in the $60^\circ$ direction by nearly 10 dB.

Fig.\ref{fig11}(a) shows the constellation diagram of the received signal at $-20^\circ$ without applying the bias voltage to the RIS, and Fig.\ref{fig11}(b) shows that when the bias voltage is applied. Similarly, the constellation diagram of the received signal is more convergent with the application of bias voltage. The spectrum of the received signal without the bias voltage is shown in Fig.\ref{fig11}(c) and that with the bias voltage is shown in Fig.\ref{fig11}(d). These results indicate that the introduction of RIS increases the RRP in the $-20^\circ$ direction by more than 20 dB.

\begin{table}[t]
	\footnotesize
	%	\vspace{-0.5cm}
	\caption{Transmission Parameters}\vspace{-1em}
	\begin{center}
		\begin{tabular}{l|l}
			\Xhline{0.7px}
			Parameters & Values or Types \\
			\hline
                Tranmit Power & -20 dBm\\
                IF Frequency & 5 GHz \\
			IFFT Points & 2048 \\
                Pilot Type & Block, ZC sequence\\
			Channel Coding & $1/2$ convolution \\
                Modulation Type & 16 QAM\\
			Baseband Sample Rate & 2.4 Gsamples/s\\
			Distance between the Transmitter and RIS & 20 cm \\
			Distance between the RIS and Receiver & 20 cm\\
			\Xhline{0.7px}
		\end{tabular}
	\end{center}\label{tab3}\vspace{0em}
\end{table}

\begin{figure}
	\centering
	\includegraphics[width=0.44\textwidth]{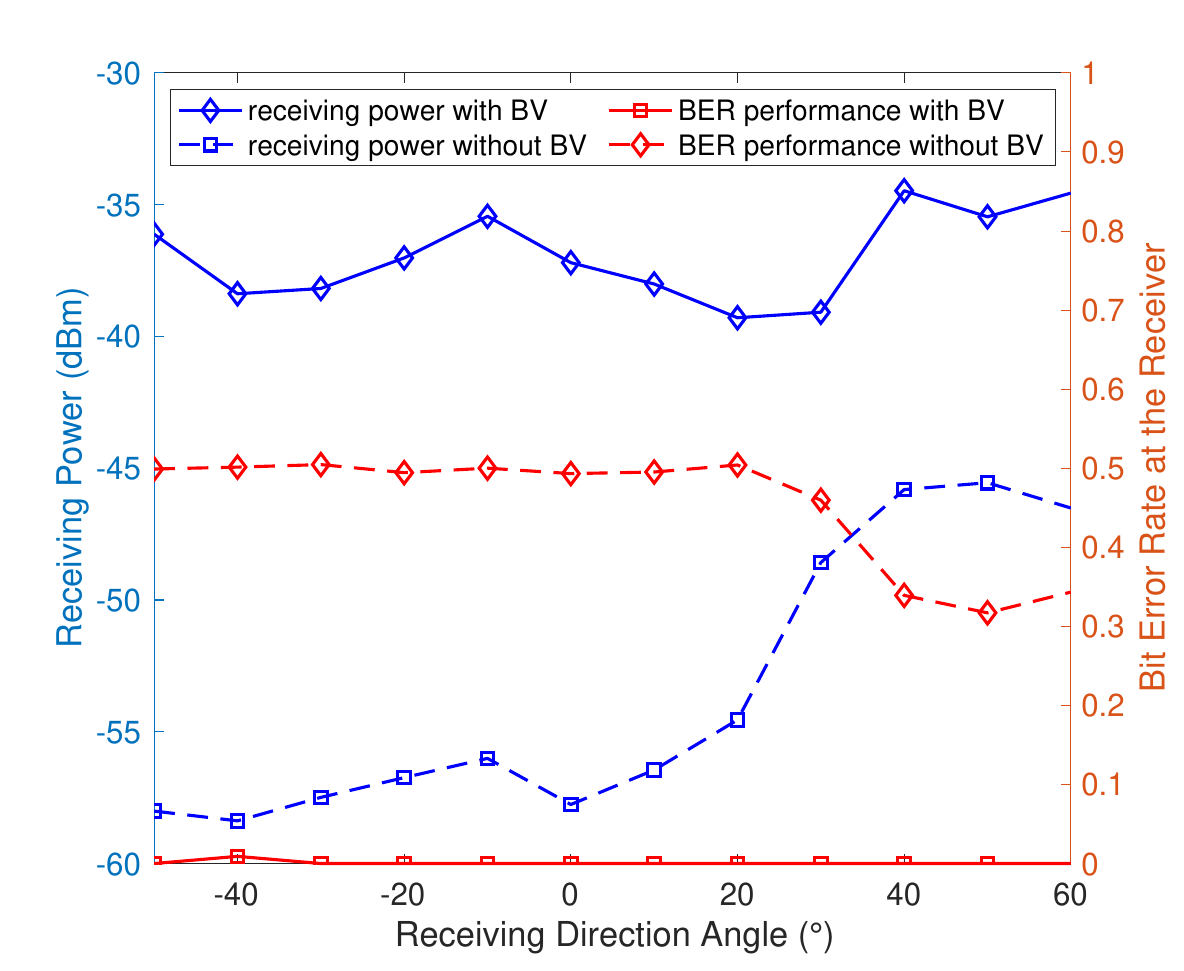}\vspace{0em}\\
	\caption{RRP and BER performance ranging from $-50^\circ$ to $60^\circ$ in the single user scenario.}\label{fig12}\vspace{0.0em}
\end{figure}

Fig.\ref{fig12} displays the performance of RRP and BER over all directions. It can be seen that when the bias voltage is applied to the RIS, the RRP in each direction exceeds -40 dBm, which is significantly improved compared to the case without the bias voltage. Specifically, in some large-angle deflection directions\footnote{In our work, the directions away from the specular reflection angle, i.e. $77^\circ$, are referred to as the large angle deflection direction.}, for example, on the $-50^\circ$ direction, the gain of RRP even exceeds 40 dB. On the other hand, the gains on small-angle deflection directions, such as $60^\circ$ and $50^\circ$, are relatively small. This can be attributed to the fact that these directions are proximate to the specular reflection direction, meaning these directions are more susceptible to the side-lobe leakage of the transmitted signal. However, in the absence of the bias voltage, even though the RRP in these small-angle directions is greater than that in the other directions, the BER of the receiver remains above 0.35. The introduction of RIS can reduce the BER to nearly 0 in most directions, highlighting the irreplaceable role of our designed RIS in THz communication systems. Notably, the effective data transmission rate\footnote{The effective data transmission rate is calculated excluding pilot and channel coding overhead.} of the designed system could reach 2.341 Gbps.

\subsection{Verification of Multi-Users Scenario}
In this subsection, we evaluated the performance of the designed system in a multi-user scenario, where the RIS could steer the incoming signal to three desired directions. The OFDM frame designed in Fig.\ref{fig7} is adopted. Each user occupies a portion of the resources in a multiplexing frame, thereby realizing an independent, interference-free transmission. The physical layer transmission parameters are the same as specified in Tab.\ref{tab3}. Due to the limited number of receivers, while keeping the voltage applied to the RIS fixed, we utilized a single receiver to sequentially receive and demodulate signals from different directions based on the resource allocation in each direction as shown in Fig.\ref{fig8}.

\begin{figure}
	\centering
	\includegraphics[width=0.48\textwidth]{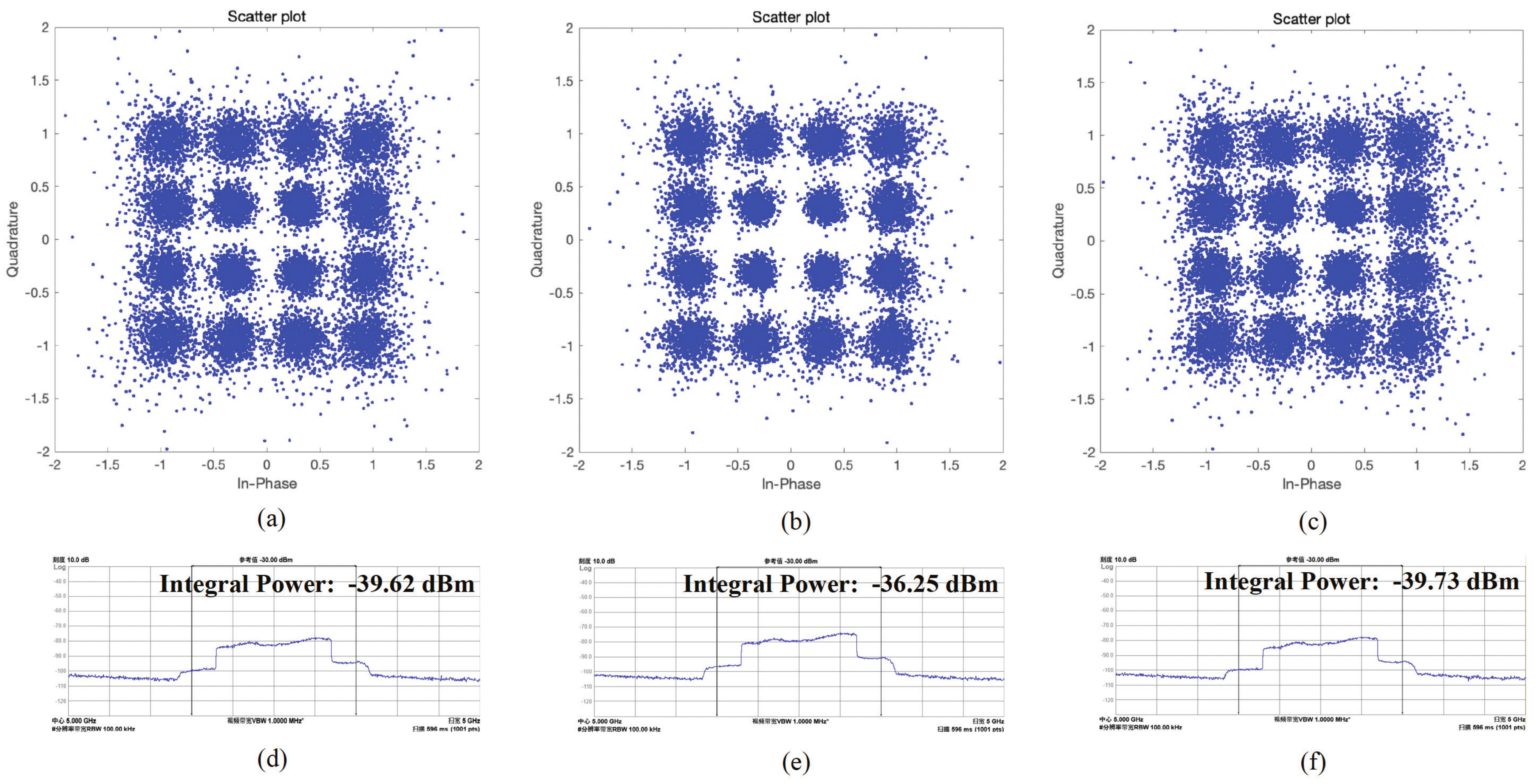}\\
	\caption{Received constellation diagrams and spectra with a bias voltage in simultaneous transmission to the $(60^\circ, 50^\circ, 40^\circ)$ directions, (a) received constellation diagram on $60^\circ$; (b) received constellation diagram on $50^\circ$; (c) received constellation diagram on $40^\circ$; (d) the received spectrum on $60^\circ$; (e) received spectrum on $50^\circ$; (f) received spectrum on $40^\circ$.}\label{fig13}
\end{figure}

Fig.\ref{fig13} exhibits the constellation diagrams and spectra of the received signal in a simultaneous transmission to the $(60^\circ, 50^\circ, 40^\circ)$ directions. It can be seen that constellation diagrams in Fig.\ref{fig13} are more converged in each direction, and the RRP on all the sub-directions in $(60^\circ, 50^\circ, 40^\circ)$ is improved significantly compared to the case without the bias voltage in Fig.\ref{fig12}. A more detailed description of RRP and BER in the multi-user scenario is shown in Fig.\ref{fig14}. Here, we illustrate RRP and BER within four groups of multi-directional beam configurations including $(-50^\circ, -20^\circ, 50^\circ)$, $(40^\circ, 50^\circ, 60^\circ)$, $(-20^\circ, 0^\circ, -20^\circ)$, and $(0^\circ, 10^\circ, 30^\circ)$. 

It can be observed that without the applied bias voltage, RRPs in most directions remain below -50 dBm, and BERs exceed 0.35. With the assistance of RIS, the RRP across all directions is improved to above -50 dBm, and the BER of the system significantly decreases. In our experiments, different information was transmitted to users in three directions. With the BER performance in Fig.\ref{fig14}, the combined peak effective data rate for all users still exceed 2 Gbps, which demonstrates the great potential of our designed RIS-THz communication system in multi-user scenarios.

\subsection{Verification of Feedback Control}
\begin{figure}
	\centering
	\includegraphics[width=0.44\textwidth]{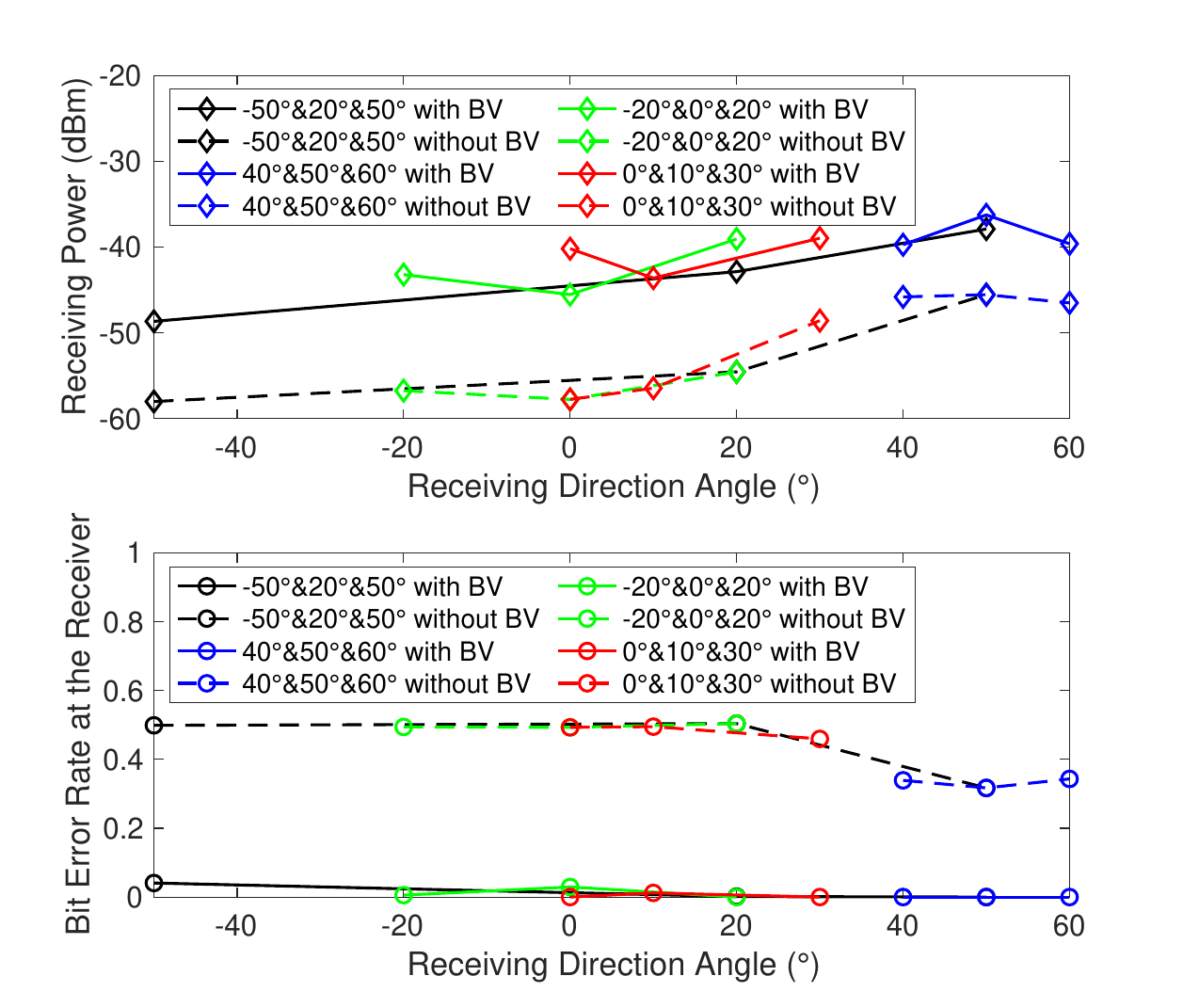}\\
	\caption{RRP and BER performance of four multi-direction user groups, including $(-50^\circ, -20^\circ, 50^\circ)$, $(40^\circ, 50^\circ, 60^\circ)$, $(-20^\circ, 0^\circ, -20^\circ)$, and $(0^\circ, 10^\circ, 30^\circ)$. }\label{fig14}
\end{figure}

An interesting fact is that, comparing the results in Fig.\ref{fig14} to that in Fig.\ref{fig12}, the enhancement effect of RIS for multi-beam scenarios appears to be smaller than that for single-beam scenarios. In Fig.\ref{fig12}, the maximum power gain exceeds 20 dB, whereas in Fig.\ref{fig14}, the power gain is no more than 15 dB. Additionally, in Fig.\ref{fig12}, the introduction of RIS could reduce the BER to zero in almost all directions, while in Fig.\ref{fig14}, although the RIS could significantly reduce the BER, there are still a few sub-directions where the BER does not drop to zero. As mentioned earlier, the optimal voltage pattern obtained for multi-beam optimization may not result in the optimal voltage pattern for each sub-direction. Although the BER in Fig.\ref{fig14} is acceptable in most scenarios, it is possible to further reduce the BER by finely adjusting the power allocation between the multiple directions.

\begin{figure*}
	\centering
	\includegraphics[width=0.85\textwidth]{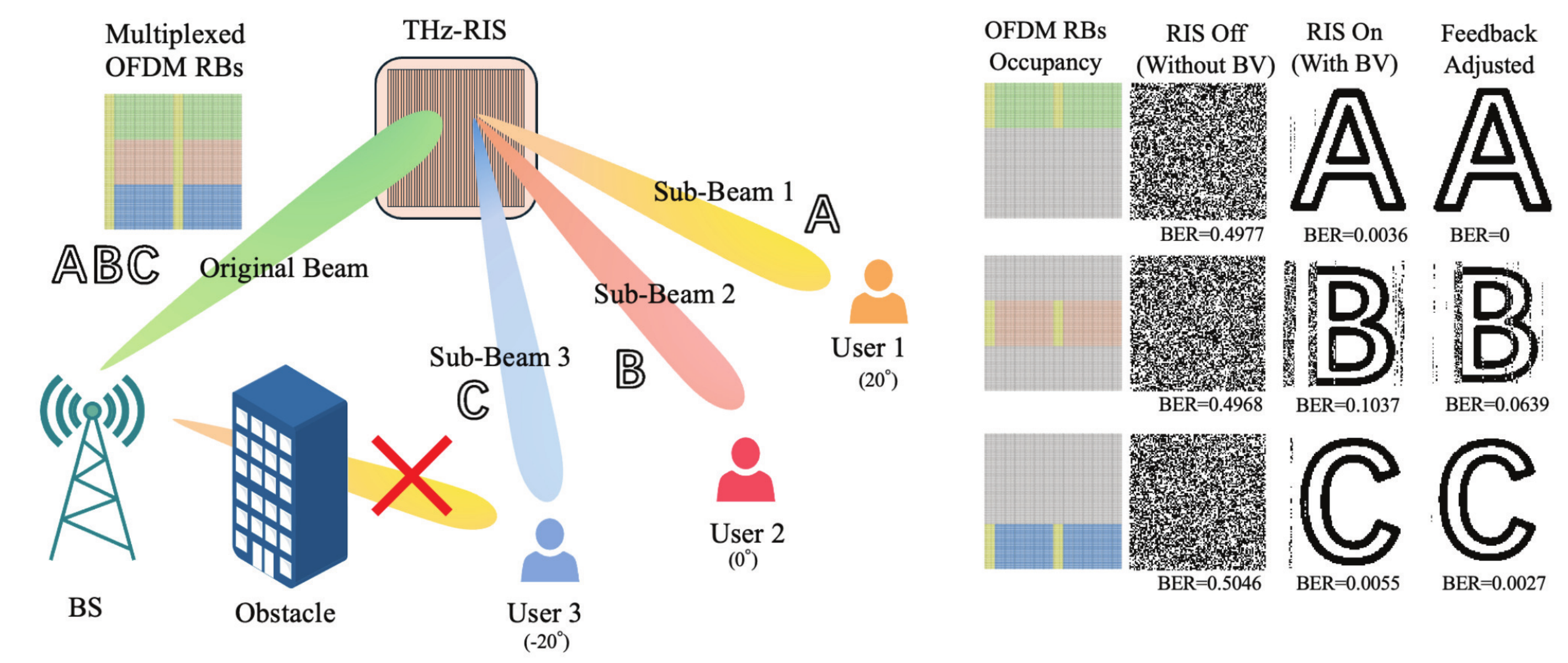}\vspace{0em}\\
	\caption{Overview of the experimental scenario, resource allocation, and experimental results with multiple users and feedback adjustments.}\label{fig15}\vspace{0.0em}
\end{figure*}

In this subsection, we validate that the designed power feedback adjustment method could further improve system performance. To vividly demonstrate the effect of power feedback adjustment, we conducted an image transmission experiment. As shown in Fig.\ref{fig15}, our objective is to simultaneously transmit three binary images, each containing a single letter A, B, and C, to three users located at $-20^\circ$, $0^\circ$, and $20^\circ$, respectively. At the transmitter, three images are first converted into bit sequences, coded, modulated, and encapsulated in a multiplexing data frame for transmission. The initial beam is split into three beams by the RIS and radiated toward three directions. Then, the user in a certain direction independently demodulates and recovers its special image based on a predefined OFDM resource allocation scheme. The physical layer transmission parameters used in this validation are the same as shown in Tab.\ref{tab3}. 

Fig.\ref{fig15} shows the scenario of the experiment, the resource allocation method, and the final experimental results. It can be seen that without applying the bias voltage to the RIS, the BER in all directions was around $50\%$, and the recovered images were chaotic and indistinguishable. After applying the bias voltage to the RIS, the BER in each direction decreased significantly, and we were able to discern the content of the received images. However, it was also evident that there were still some bit errors in each direction, manifested as unexpected black pixels or missing pixels in the images. Notably, the image quality in the $0^\circ$ direction was worse than in the other two directions. 

To improve the quality of received images for all the 3 users, we applied the proposed power feedback adjustment method to fine-tune the beams for multiple sub-directions. After the feedback adjustment, the image quality recovered by users in all directions was enhanced. The BER on the $20^\circ$ and $-20^\circ$ directions reduced to 0 and 0.0027, with almost no erroneous pixels in the recovered images. Although the BER on the $0^\circ$ direction did not reach 0, it's also improved significantly, with noticeably fewer erroneous pixels compared to the images received before feedback adjustment. This demonstrates that the power feedback adjustment method can further improve the quality of received signals. In practical applications, whether operating the power feedback adjustment can be decided on the actual performance of the system after the bias voltage is applied.

\section{Conclusion}\label{VI}
In this paper, we first introduce the structure, frequency characteristics, and control methods of our designed 220 GHz RIS, which is capable of continuously adjusting the deflection of the incident beam over a range of $110^\circ$, and steering the highly focused THz beams towards multiple directions. Secondly, we design a multi-user-oriented THz physical layer transmission scheme, including the design of a multiplexing scheme and the prototype of the hardware system. The THz communication system could achieve Gbps-level data transmission rates. Furthermore, we introduce an SPGD-based beamforming method and a power feedback adjustment method for the designed RIS, which are used to roughly and finely adjust the quality of the received signal in the desired directions.

Eventually, we built a multi-user RIS-THz communication system based on the designed devices and methods. To the best of our knowledge, this is the first attempt to integrate RIS into the practical THz band multi-user communication systems. We perform beamforming in the desired directions, including single-direction beam deflection and multi-direction beam splitting. We measured the RRP in both single-user and multi-user scenarios and calculated the BER of the system. In the single-user scenario, the introduction of the RIS provided a power gain of no less than 10 dB for the THz communication system, while in the multi-user scenario, the gain was no less than 5 dB. Finally, through an image transmission experiment, we demonstrated that the designed power feedback adjustment method can fine-tune the quality of the received signal for each user in a multi-user scenario. Our experimental results indicate that RIS has great potential for future THz communication systems.

%\section*{Acknowledgments}
%This should be a simple paragraph before the References to thank those individuals and institutions who have supported your work on this article.

\bibliographystyle{IEEEtran}
\bibliography{TCOMM}

\begin{thebibliography}{10}
\providecommand{\url}[1]{#1}
\csname url@samestyle\endcsname
\providecommand{\newblock}{\relax}
\providecommand{\bibinfo}[2]{#2}
\providecommand{\BIBentrySTDinterwordspacing}{\spaceskip=0pt\relax}
\providecommand{\BIBentryALTinterwordstretchfactor}{4}
\providecommand{\BIBentryALTinterwordspacing}{\spaceskip=\fontdimen2\font plus
\BIBentryALTinterwordstretchfactor\fontdimen3\font minus
  \fontdimen4\font\relax}
\providecommand{\BIBforeignlanguage}[2]{{%
\expandafter\ifx\csname l@#1\endcsname\relax
\typeout{** WARNING: IEEEtran.bst: No hyphenation pattern has been}%
\typeout{** loaded for the language `#1'. Using the pattern for}%
\typeout{** the default language instead.}%
\else
\language=\csname l@#1\endcsname
\fi
#2}}
\providecommand{\BIBdecl}{\relax}
\BIBdecl

\bibitem{saad2019vision}
W.~Saad, M.~Bennis, and M.~Chen, ``A vision of 6g wireless systems:
  Applications, trends, technologies, and open research problems,'' \emph{IEEE
  Netw.}, vol.~34, no.~3, pp. 134--142, 2019.

\bibitem{akyildiz20206g}
I.~F. Akyildiz, A.~Kak, and S.~Nie, ``6g and beyond: The future of wireless
  communications systems,'' \emph{IEEE Access}, vol.~8, pp. 133\,995--134\,030,
  2020.

\bibitem{yang20196g}
P.~Yang, Y.~Xiao, M.~Xiao, and S.~Li, ``6g wireless communications: Vision and
  potential techniques,'' \emph{IEEE Netw.}, vol.~33, no.~4, pp. 70--75, 2019.

\bibitem{ji2021several}
B.~Ji, Y.~Han, S.~Liu, F.~Tao, G.~Zhang, Z.~Fu, and C.~Li, ``Several key
  technologies for 6g: Challenges and opportunities,'' \emph{IEEE Commun.
  Standards Mag.}, vol.~5, no.~2, pp. 44--51, 2021.

\bibitem{chafii2023twelve}
M.~Chafii, L.~Bariah, S.~Muhaidat, and M.~Debbah, ``Twelve scientific
  challenges for 6g: Rethinking the foundations of communications theory,''
  \emph{IEEE Commun. Surveys Tuts.}, vol.~25, no.~2, pp. 868--904, 2023.

\bibitem{yuan2020potential}
Y.~Yuan, Y.~Zhao, B.~Zong, and S.~Parolari, ``Potential key technologies for 6g
  mobile communications,'' \emph{Sci. China Inf. Sci.}, vol.~63, no.~8, p.
  183301, 2020.

\bibitem{shafie2022terahertz}
A.~Shafie, N.~Yang, C.~Han, J.~M. Jornet, M.~Juntti, and T.~K{\"u}rner,
  ``Terahertz communications for 6g and beyond wireless networks: Challenges,
  key advancements, and opportunities,'' \emph{IEEE Netw.}, vol.~37, no.~3, pp.
  162--169, 2022.

\bibitem{alsharif2021toward}
M.~H. Alsharif, M.~A. Albreem, A.~A.~A. Solyman, and S.~Kim, ``Toward 6g
  communication networks: Terahertz frequency challenges and open research
  issues,'' \emph{Comput. Mater. Continua}, vol.~66, no.~3, pp. 2831--2842,
  2021.

\bibitem{chowdhury20206g}
M.~Z. Chowdhury, M.~Shahjalal, S.~Ahmed, and Y.~M. Jang, ``6g wireless
  communication systems: Applications, requirements, technologies, challenges,
  and research directions,'' \emph{IEEE Open J. Commun. Soc.}, vol.~1, pp.
  957--975, 2020.

\bibitem{yang2022terahertz}
N.~Yang and A.~Shafie, ``Terahertz communications for massive connectivity and
  security in 6g and beyond era,'' \emph{IEEE Commun. Mag.}, vol.~62, no.~2,
  pp. 72--78, 2022.

\bibitem{10255735}
Y.~Chen, J.~Tan, M.~Hao, R.~MacKenzie, and L.~Dai, ``Accurate beam training for
  ris-assisted wideband terahertz communication,'' \emph{IEEE Trans. Commun.},
  vol.~71, no.~12, pp. 7425--7440, 2023.

\bibitem{song2021terahertz}
H.-J. Song and N.~Lee, ``Terahertz communications: Challenges in the next
  decade,'' \emph{IEEE Trans. THz Sci. Technol.}, vol.~12, no.~2, pp. 105--117,
  2021.

\bibitem{chataut2020massive}
R.~Chataut and R.~Akl, ``Massive mimo systems for 5g and beyond
  networks—overview, recent trends, challenges, and future research
  direction,'' \emph{Sensors}, vol.~20, no.~10, p. 2753, 2020.

\bibitem{do2021terahertz}
H.~Do, S.~Cho, J.~Park, H.-J. Song, N.~Lee, and A.~Lozano, ``Terahertz
  line-of-sight mimo communication: Theory and practical challenges,''
  \emph{IEEE Commun. Mag.}, vol.~59, no.~3, pp. 104--109, 2021.

\bibitem{liu2021reconfigurable}
Y.~Liu, X.~Liu, X.~Mu, T.~Hou, J.~Xu, M.~Di~Renzo, and N.~Al-Dhahir,
  ``Reconfigurable intelligent surfaces: Principles and opportunities,''
  \emph{IEEE Commun. Surveys Tuts.}, vol.~23, no.~3, pp. 1546--1577, 2021.

\bibitem{yuan2021reconfigurable}
X.~Yuan, Y.-J.~A. Zhang, Y.~Shi, W.~Yan, and H.~Liu,
  ``Reconfigurable-intelligent-surface empowered wireless communications:
  Challenges and opportunities,'' \emph{IEEE Wireless Commun.}, vol.~28, no.~2,
  pp. 136--143, 2021.

\bibitem{pan2021reconfigurable}
C.~Pan, H.~Ren, K.~Wang, J.~F. Kolb, M.~Elkashlan, M.~Chen, M.~Di~Renzo,
  Y.~Hao, J.~Wang, A.~L. Swindlehurst \emph{et~al.}, ``Reconfigurable
  intelligent surfaces for 6g systems: Principles, applications, and research
  directions,'' \emph{IEEE Commun. Mag.}, vol.~59, no.~6, pp. 14--20, 2021.

\bibitem{basharat2021reconfigurable}
S.~Basharat, S.~A. Hassan, H.~Pervaiz, A.~Mahmood, Z.~Ding, and M.~Gidlund,
  ``Reconfigurable intelligent surfaces: Potentials, applications, and
  challenges for 6g wireless networks,'' \emph{IEEE Wireless Commun.}, vol.~28,
  no.~6, pp. 184--191, 2021.

\bibitem{10720781}
M.~Ahmed, A.~Wahid, W.~U. Khan, F.~Khan, A.~Ihsan, Z.~Ali, K.~Rabie,
  T.~Shongwe, and Z.~Han, ``A survey on ris advances in terahertz
  communications: Emerging paradigms and research frontiers,'' \emph{IEEE
  Access}, pp. 1--1, 2024.

\bibitem{raza2022intelligent}
A.~Raza, U.~Ijaz, M.~K. Ishfaq, S.~Ahmad, M.~Liaqat, F.~Anwar, A.~Iqbal, and
  M.~S. Sharif, ``Intelligent reflecting surface-assisted terahertz
  communication towards b5g and 6g: State-of-the-art,'' \emph{Microw. Opt.
  Technol. Lett.}, vol.~64, no.~5, pp. 858--866, 2022.

\bibitem{yang2022terahertz2}
F.~Yang, P.~Pitchappa, and N.~Wang, ``Terahertz reconfigurable intelligent
  surfaces (riss) for 6g communication links,'' \emph{Micromachines}, vol.~13,
  no.~2, p. 285, 2022.

\bibitem{huang2021multi}
C.~Huang, Z.~Yang, G.~C. Alexandropoulos, K.~Xiong, L.~Wei, C.~Yuen, Z.~Zhang,
  and M.~Debbah, ``Multi-hop ris-empowered terahertz communications: A
  drl-based hybrid beamforming design,'' \emph{IEEE J. Sel. Areas Commun.},
  vol.~39, no.~6, pp. 1663--1677, 2021.

\bibitem{do2024throughput}
T.~Do-Duy, A.~Masaracchia, B.~Canberk, L.~D. Nguyen, and T.~Q. Duong,
  ``Throughput maximisation in ris-assisted noma-thz communication network,''
  \emph{IEEE Open J. Commun. Soc.}, vol.~5, pp. 5706--5717, 2024.

\bibitem{dai2019wireless}
J.~Y. Dai, W.~K. Tang, J.~Zhao, X.~Li, Q.~Cheng, J.~C. Ke, M.~Z. Chen, S.~Jin,
  and T.~J. Cui, ``Wireless communications through a simplified architecture
  based on time-domain digital coding metasurface,'' \emph{Adv. Mater.
  Technol.}, vol.~4, no.~7, p. 1900044, 2019.

\bibitem{liu2023toward}
Y.~Liu, Y.~Wang, X.~Fu, L.~Shi, F.~Yang, J.~Luo, Q.~Y. Zhou, Y.~Fu, Q.~Chen,
  J.~Y. Dai \emph{et~al.}, ``Toward sub-terahertz: Space-time coding
  metasurface transmitter for wideband wireless communications,'' \emph{Adv.
  Sci.}, vol.~10, no.~29, p. 2304278, 2023.

\bibitem{wu2023universal}
G.-B. Wu, J.~Y. Dai, K.~M. Shum, K.~F. Chan, Q.~Cheng, T.~J. Cui, and C.~H.
  Chan, ``A universal metasurface antenna to manipulate all fundamental
  characteristics of electromagnetic waves,'' \emph{Nat. Commun.}, vol.~14,
  no.~1, p. 5155, 2023.

\bibitem{wang2022broadband}
W.~Wang, E.~Lv, Y.~Hou, and D.~Yang, ``Broadband beam steering based on
  programmable vo 2 metasurface at terahertz frequencies,'' in \emph{2022 Asia
  Communications and Photonics Conference (ACP)}.\hskip 1em plus 0.5em minus
  0.4em\relax IEEE, 2022, pp. 50--52.

\bibitem{lv2023broadband}
E.~Lv, W.~Wang, S.~Liu, T.~Pan, C.~Zhu, H.~Wang, B.~Liu, Y.~Hou, and D.~Yang,
  ``Broadband programmable metasurface for multifunctional control of thz
  waves,'' in \emph{CLEO: Fundamental Science}.\hskip 1em plus 0.5em minus
  0.4em\relax Optica Publishing Group, 2023, pp. JTh2A--110.

\bibitem{wu2020liquid}
J.~Wu, Z.~Shen, S.~Ge, B.~Chen, Z.~Shen, T.~Wang, C.~Zhang, W.~Hu, K.~Fan,
  W.~Padilla \emph{et~al.}, ``Liquid crystal programmable metasurface for
  terahertz beam steering,'' \emph{Appl. Phys. Lett.}, vol. 116, no.~13, 2020.

\bibitem{li2023modulo}
W.~Li, B.~Chen, X.~Hu, H.~Guo, S.~Wang, J.~Wu, K.~Fan, C.~Zhang, H.~Wang,
  B.~Jin \emph{et~al.}, ``Modulo-addition operation enables terahertz
  programmable metasurface for high-resolution two-dimensional beam steering,''
  \emph{Sci. Adv.}, vol.~9, no.~42, p. eadi7565, 2023.

\bibitem{chen2024liquid}
C.~Chen, S.~Chen, Y.~Ni, Y.~Xu, and Y.~Yang, ``Liquid crystal metasurface for
  on-demand terahertz beam forming over 110° field-of-view,'' \emph{Laser
  Photonics Rev.}, p. 2400237, 2024.

\bibitem{10177872}
J.~Tang, M.~Cui, S.~Xu, L.~Dai, F.~Yang, and M.~Li, ``Transmissive ris for b5g
  communications: Design, prototyping, and experimental demonstrations,''
  \emph{IEEE Trans. Commun.}, vol.~71, no.~11, pp. 6605--6615, 2023.

\bibitem{liu2018programmable}
F.~Liu, A.~Pitilakis, M.~S. Mirmoosa, O.~Tsilipakos, X.~Wang, A.~C.
  Tasolamprou, S.~Abadal, A.~Cabellos-Aparicio, E.~Alarc{\'o}n, C.~Liaskos
  \emph{et~al.}, ``Programmable metasurfaces: State of the art and prospects,''
  in \emph{2018 IEEE International Symposium on Circuits and Systems
  (ISCAS)}.\hskip 1em plus 0.5em minus 0.4em\relax IEEE, 2018, pp. 1--5.

\bibitem{shafie2021spectrum}
A.~Shafie, N.~Yang, S.~A. Alvi, C.~Han, S.~Durrani, and J.~M. Jornet,
  ``Spectrum allocation with adaptive sub-band bandwidth for terahertz
  communication systems,'' \emph{IEEE Trans. Commun.}, vol.~70, no.~2, pp.
  1407--1422, 2021.

\bibitem{yan2023beamforming}
W.~Yan, W.~Hao, C.~Huang, G.~Sun, O.~Muta, H.~Gacanin, and C.~Yuen,
  ``Beamforming analysis and design for wideband thz reconfigurable intelligent
  surface communications,'' \emph{IEEE J. Sel. Areas Commun.}, vol.~41, no.~8,
  pp. 2306--2320, 2023.

\bibitem{10184278}
W.~Hao, H.~Shi, G.~Sun, and C.~Huang, ``Joint beamforming design for active
  ris-aided thz isac systems with delay alignment modulation,'' \emph{IEEE
  Wireless Commun. Lett.}, vol.~12, no.~10, pp. 1816--1820, 2023.

\bibitem{10684280}
I.~Yildirim, A.~Koc, E.~Basar, and T.~Le-Ngoc, ``Multi-ris assisted hybrid
  beamforming design for terahertz massive mimo systems,'' \emph{IEEE Open J.
  Commun. Soc.}, vol.~5, pp. 6150--6165, 2024.

\bibitem{vorontsov1998stochastic}
M.~A. Vorontsov and V.~P. Sivokon, ``Stochastic parallel-gradient-descent
  technique for high-resolution wave-front phase-distortion correction,''
  \emph{J. Opt. Soc. Am.}, vol.~15, no.~10, pp. 2745--2758, 1998.

\end{thebibliography}

% your own custom macro containing the $\backslash${\tt{includegraphics}} command to make things
 
%\vspace{-33pt}
%\begin{IEEEbiography}[{\includegraphics[width=1in,height=1.25in,clip,keepaspectratio]{fig1}}]{Michael Shell}
%Use $\backslash${\tt{begin\{IEEEbiography\}}} and then for the 1st argument use $\backslash${\tt{includegraphics}} to declare and link the author photo.
%Use the author name as the 3rd argument followed by the biography text.
%\end{IEEEbiography}
%\vspace{-33pt}

\vfill
\end{document}